\documentclass[showpacs,preprintnumbers,amsmath,amssymb]{revtex4}

\usepackage{graphicx}
\usepackage{dcolumn}
\usepackage{bm}

\def\be{\begin{equation}}
\def\ee{\end{equation}}
\def\bea{\begin{eqnarray}}
\def\eea{\end{eqnarray}}
\def\NO{\nonumber}
\def\gev{\mathrm{~GeV}}

\begin{document}


\title{Global analysis of the experimental data on $\chi_c$ meson hadroproduction}


\author{Hong-Fei Zhang$^{1,3}$}
\email{hfzhang@ihep.ac.cn}
\author{Ling Yu$^{2}$}
\author{Shao-Xiang Zhang$^{3}$}
\author{Lan Jia$^{1}$}
\affiliation{
$^{1}$ Department of Physics, School of Biomedical Engineering, Third Military Medical University, Chongqing 400038, China\\
$^{2}$ Department of Power Engineering, Wuhan Electric Technical Power College, Wuhan 430079, China\\
$^{3}$ Department of Digital Medical Science, School of Biomedical Engineering, Third Military Medical University, Chongqing 400038, China
}%
\date{\today}

\begin{abstract}
We carry out a global analysis of the experimental data on the $\chi_c$ production cross section and the ratio $\sigma(\chi_{c2})/\sigma(\chi_{c1})$ at the LHC and the Tevatron.
The related long-distance matrix elements (LDMEs) at both leading order (LO) and next-to-leading order (NLO) in the QCD coupling constant are renewed.
We also present the transverse momentum distribution of the $\chi_c$ production cross section and the ratio $\sigma(\chi_{c2})/\sigma(\chi_{c1})$ for several experimental conditions
and find that NLO predictions agree with all sets of experimental data.
By contrast, at LO, one cannot explain all the data with a unique value of the color-octet LDME.
A brief analysis of the nonrelativistic QCD scale dependence of the cross sections shows that,
for the conditions we are concerned with in this paper, the dependence can be almost totally absorbed into the LDME.
\end{abstract}

\maketitle
\section{introduction}
Since the Large Hadron Collider (LHC) started its run, many experimental results have come out
that have provided an opportunity to carry out further investigation of the phenomenology of QCD-based effective theories.
Nonrelativistic QCD (NRQCD) is one of the most successful effective theories describing quarkonium production and decays~\cite{Bodwin:1994jh}.
Under the NRQCD framework, the cross section is factorized into the summation of the products of the short-distance coefficient (SDC),
which is independent of the quarkonium state and can be calculated perturbatively, and the long-distance matrix element (LDME),
which only depends on the quarkonium state and requires the fit of experimental data to extract its value.
The cross section for the process of heavy quarkonium $H$ production or decay can be expressed as
\be
d\sigma(H)=\sum_ndf_n\langle O^H(n)\rangle \label{eqn:NRQCD}
\ee
where $f_n$ is the SDC for the $q\bar{q}$ state $n$, and $\langle O^H(n)\rangle$ is the LDME of state $n$ for quarkonium $H$.

NRQCD succeeded in many processes where the color-singlet (CS) model ~\cite{Einhorn:1975ua, Ellis:1976fj, Chang:1979nn, Berger:1980ni, Baier:1981zz} failed;
however, it still faces many challenges.
$J/\psi$ polarization at hadron colliders is among the most puzzling questions that NRQCD encounters.
References~\cite{Butenschoen:2012px, Chao:2012iv} investigated the polarization of directly produced $J/\psi$,
while Ref.~\cite{Gong:2012ug} provided polarization results for prompt $J/\psi$ hadroproduction,
which is the first next-to-leading-order (NLO) result comparable with experiment.
The three letters employed three sets of LDMEs, which were obtained from different fit strategies.
All of them can describe $J/\psi$ production, yet
none of them can explain all the polarization measurements.
In addition, the universality of the LDMEs is another challenge.
Reference~\cite{Butenschoen:2010rq} reconciled experimental data of $J/\psi$ production at the Tevatron and HERA;however,
their LDMEs resulted in unphysical cross sections when employed to $J/\psi$ associated with a photon production at hadron colliders~\cite{Li:2014ava}.
Another interesting example is the transverse momentum ($p_t$) integrated cross sections for the $J/\psi$ hadroproduction.
The theoretical results at QCD NLO obtained with collinear factorization~\cite{Maltoni:2006yp, Feng:2015cba} overshoot the experimental data.
Having resummed $log(x)$ (where $x$ denotes the Bjorken-$x$) and considered the all-twist contributions in the dense side, Ma and Venugopalan~\cite{Ma:2014mri} remedied the discrepancy.
However, they~\cite{Ma:2014mri} did not include the $\chi_c$ and $\psi'$ feeddown contributions.
Whether the inclusion of these parts will ruin the conclusions requires further investigation.
Actually, Ref.~\cite{Hagler:2000dd} has already studied the $\chi_c$ production processes in which the $p_t$ of the $\chi_c$ completely comes from the initial states.
Exploiting the unintegrated gluon distribution,
the authors announced that the color-octet (CO) LDME for $\chi_c$ is almost an order of magnitude smaller than the one obtained through the collinear factorization calculations.
Reference~\cite{Shuvaev:2015fta}, however, found that the final-state gluon emission processes actually dominate the $\chi_c$ hadroproduction in the midrapidity region.
For the above reasons, testing NRQCD is still an important work.

P-wave quarkonia productions provide an excellent laboratory to test NRQCD.
Above all, at LO in $v$ (the typical relative velocity of quark and antiquark in quarkonium),
only one LDME is to be obtained from experiment;
one does not suffer from the entanglement of too many free parameters in the fit of experimental data,
 in contrast to the $J/\psi$ case~\cite{Butenschoen:2010rq, Butenschoen:2012px, Ma:2010yw, Ma:2010jj, Chao:2012iv, Gong:2012ug}.
Further, at NLO, the $p_t$ distribution of both $^3P_J^{[1]}$ and $^3S_1^{[8]}$ channels
behaves as $1/p_t^4$ in the large $p_t$ region.
Higher order corrections cannot exceed this behavior; thus,
one can expect NLO predictions to give a good precision in this phase space region.
 In contrast to the $^3S_1^{[1]}$ case, the significance of NNLO correction is still in the mist.
Finally, feeddown from higher excited states to P-wave quarkonia, say, $\chi_c$, $h_c$, $\chi_b$, or $h_b$, can almost be neglected
[e.g.,$\sigma(\psi(2s)\rightarrow\chi_{c1}(1p))/\sigma(\chi_{c1}(1p))\sim 5\%$ at LHCb~\cite{Aaij:2012ag, LHCb:2012af, Aaij:2011jh}].
Notice the advantages stated above:we say the case of $\chi_c$ is ``clean";
it is much easier to make definite conclusions in this case than in the $J/\psi$ case.

On the other hand, the study of $\chi_c$ production is not only important itself for phenomenological concerns,
but also provides an opportunity for precise study of $J/\psi$ phenomenology
(e.g.,Ref.~\cite{Gong:2012ug} indicates that $\chi_c$ feeddown contribution is essential to the study of $J/\psi$ polarization).
Many theoretical works on $\chi_c$ production have been published.
The authors of Refs.~\cite{Cho:1995vh, Cho:1995ce, Braaten:1999qk, Sharma:2012dy} obtained LDME for $\chi_c$ production at LO
and employed them to the prediction of prompt $J/\psi$ hadroproduction and/or polarization.
Ma \emph{et al}.~\cite{Ma:2010vd} presented the first NLO study of $\chi_c$ hadroproduction and gave a favorable choice of the LDME for $\chi_c$ production.
Li \emph{et al}.~\cite{Li:2011yc} calculated $\chi_c$ production associated with a $c\bar{c}$ pair at hadron colliders.
Shao \emph{et al}.~\cite{Shao:2012fs, Shao:2014fca} provided a detailed study on the polarization of hadroproduced $\chi_c$ and $\chi_c$-generated $J/\psi$;
at the same time,
they compared the theoretical prediction with some of the recent experimental data~\cite{Abulencia:2007bra, LHCb:2012ac, Chatrchyan:2012ub}.
Likhoded \emph{et al}.~\cite{Likhoded:2014kfa} calculated $\chi_c$ hadroproduction at LO in $\alpha_s$
and extracted both the CS and CO LDMEs from the fit of the experimental data,
where their CS LDME is several times larger than the value obtained by the potential model
and higher order terms in $v^2$ contribute significantly.

This paper is devoted to the theoretical predictions of $\chi_c$ hadroproduction, for one thing, as an alternative test of NRQCD.
Recently, a number of experiment results have come out from the LHC collaborations,
among which are many measurements on the $\chi_c$ yield and the ratio $\sigma(\chi_{c2})/\sigma(\chi_{c1})$.
This paper will answer the question whether a single LDME can explain all the experimental data.
For another thing, the popular values of the LDMEs for $\chi_c$ production, both at LO and NLO,
were all given before these experimental results were published; they are out of date.
This paper will provide a detailed analysis on the determination of the LDMEs and reasonable values of it at both LO and NLO.
Finally, as suggested in Ref.~\cite{Wang:2014vsa},
we also observe the NRQCD scale dependence of the cross sections to determine whether NLO prediction stands up.

This paper is organized as follows.
Section II gives a brief introduction to the NRQCD framework for $\chi_c$ hadroproduction calculation.
In Sec. III, we present the parameter choices in our numerical computation and an analysis to see whether the NRQCD scale dependence is severe.
In Sec. IV, we present the results of the fit and the values of the LDMEs at both LO and NLO and investigate the universality of the LDMEs in detail.
Section VI is a concluding remark.

\section{$\chi_c$ production in NRQCD framework}

This section provides quite a brief review of the NRQCD formulas for the calculation of $\chi_c$ production.
We do not discuss in detail how the equations are derived.
Interested readers can refer to some relative references, e.g. ~\cite{Petrelli:1997ge, Wang:2014vsa}.

For $\chi_c$ production at LO in $v$, Eq.(\ref{eqn:NRQCD}) can be written as
\be
d\sigma(\chi_{cJ})=df_{^3P_J^{[1]}}\langle O^{\chi_{cJ}}(^3P_J^{[1]})\rangle
+df_{^3S_1^{[8]}}(2J+1)\langle O^{\chi_{c0}}(^3S_1^{[8]})\rangle \label{eqn:s}
\ee.
The value of CS LDME can be evaluated through~\cite{Bodwin:1994jh}
\be
\langle O^{\chi_{cJ}}(^3P_J^{[1]})\rangle=\frac{9}{2\pi}(2J+1)|R_p'(0)|^2
\ee,
where $R_p'(0)$ is the derivative of the wave function of the related quarkonium with respect to the radius at the origin.

The calculation of $f_{^3S_1^{[8]}}$ has been described in detail in many of our previous papers;see,e.g., ~\cite{Gong:2010bk}.
To evaluate $f_{^3P_J^{[1]}}$, we notice that it is independent of the long-distance asymptotic states
and replace $\chi_{cJ}$ in Eq.(\ref{eqn:s}) with a $c\bar{c}$ state $^3P_J^{[1]}$, so that we obtain
\bea
d\sigma(^3P_J^{[1]})=df_{^3P_J^{[1]}}\langle O^{^3P_J^{[1]}}(^3P_J^{[1]})\rangle
+df_{^3S_1^{[8]}}\langle O^{^3P_J^{[1]}}(^3S_1^{[8]})\rangle \label{eqn:fac}
\eea,
where we have used the relation $\langle O^{^3P_J^{[1]}}(^3S_1^{[8]})\rangle=(2J+1)\langle O^{^3P_0^{[1]}}(^3S_1^{[8]})\rangle$.
We should keep in mind that Eq.(\ref{eqn:fac}) is to extract CS SDC, which is expanded in $\alpha_s$.
As a result, the quantities $\langle O^{^3P_J^{[1]}}(^3P_J^{[1]})\rangle$ and $\langle O^{^3P_J^{[1]}}(^{3}S_{1}^{[8]})\rangle$
should also be evaluated perturbatively, and the value of $\alpha_s$ in them should be in accordance with that in the SDCs.
The left- and right-hand side of Eq.(\ref{eqn:fac}) should keep those terms up to the same order as in the perturbative expansion.
The evaluation of $d\sigma_{^3P_J^{[1]}}$ follows the ordinary procedure:
writing the squared amplitudes through reading Feynman diagrams and multiplying it by the flux density and the phase space unit.
Both $d\sigma_{^3P_J^{[1]}}$ and $\langle O^{^3P_J^{[1]}}(^3S_1^{[8]})\rangle$ are infrared (IR) divergent,
and the IR divergences from the two quantities cancel each other.

\begin{figure}
\center{
\includegraphics*[scale=0.5]{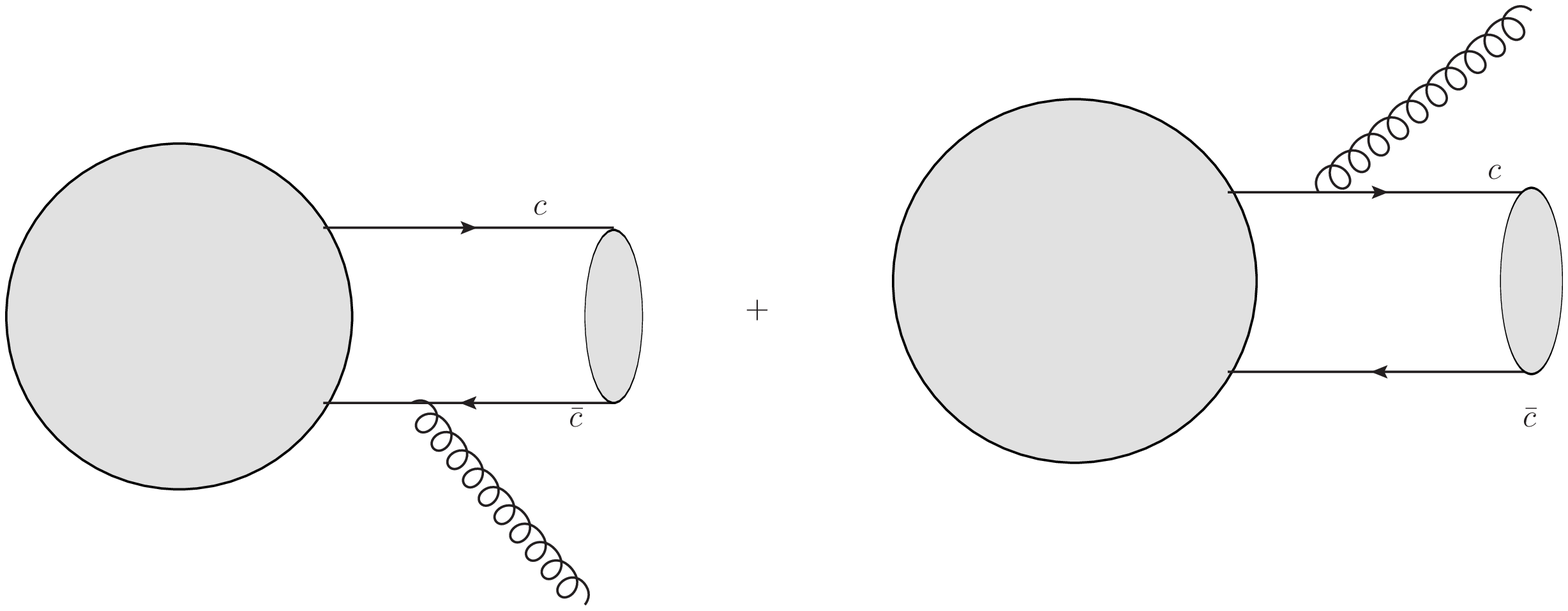}
\caption{\label{fig:gc} Typical diagrams where the soft gluon connects to the quarkonium.}
}
\end{figure}

We adopt the two-cutoff phase space slicing method~\cite{Harris:2001sx} and find that
the cross section excluding the terms (denoted as $d\sigma^S$) corresponding to the squared real-correction diagrams,
in which a gluon connects the quarkonium (as displayed in Fig.~\ref{fig:gc}),
integrated over the gluon soft region is free of divergence.
We denote this finite part of the cross section as $d\sigma^F$.
Neglecting the finite terms proportional to the size of the small region, $d\sigma^S$ can be expressed as
\be
d\sigma^S(^3P_J^{[1]})=-\frac{\alpha_s}{3\pi m_{c}^{2}}u_{\epsilon}^{s}\frac{N_{c}^{2}-1}{N_{c}^{2}}
d f_{^3S_1^{[8]}}\langle O^{^3P_J^{[1]}}(^3P_J^{[1]})\rangle , \label{eqn:scs}
\ee
where $N_c$ is 3 for SU(3) gauge field and
\be
u_{\epsilon}^{s}=\frac{1}{\epsilon_{IR}}+\frac{E}{p}ln(\frac{E+p}{E-p})+ln(\frac{4\pi\mu^{2}}{s\delta_{s}^{2}})-\gamma_{E}-\frac{1}{3} \label{eqn:us},
\ee
with $E$ and $p$ being the energy and absolute value of momentum of $\chi_c$, respectively,
$\gamma_E$ the Euler's constant, and $\mu$ the scale to complement the dimension.
$\delta_s$ is an arbitrary positive number small enough to provide the soft approximation with sufficient accuracy.

Up to the order maintained in our calculation,
the transition rate of $c\bar{c}$ state $^3S_1^{[8]}$ into $^3P_J^{[1]}$ can be calculated in the dimensional regularization scheme as
\be
\langle O^{^3P_J^{[1]}}(^3S_1^{[8]})\rangle^{NLO}=-\frac{\alpha_s}{3\pi m_{c}^2}u^{c}_{\epsilon}\frac{N_{c}^{2}-1}{N_{c}^{2}}
\langle O^{^3P_J^{[1]}}(^3P_J^{[1]})\rangle^{LO} , \label{eqn:LDM}
\ee
where $u^c_\epsilon$ is defined as
\be
u^{c}_{\epsilon}|_{\mu_\Lambda}=\frac{1}{\epsilon_{IR}}-\gamma_{E}-\frac{1}{3}+ln(\frac{4\pi\mu^2}{\mu^2_{\Lambda}}) \label{eqn:ucmu}
\ee
and
\be
u^{c}_{\epsilon}|_{\overline{MS}}=\frac{1}{\epsilon_{IR}}-\gamma_{E}+\frac{5}{3}+ln(\frac{\pi\mu^2}{\mu^2_{\Lambda}}) \label{eqn:ucms}
\ee
in the $\mu_\Lambda$-cutoff (in which $\mu_\Lambda$ is the upper bound of the integrated gluon energy)
and $\overline{MS}$ renormalization scheme, respectively.
$\mu_\Lambda$ is a scale rising from the renormalization of the LDME.

Substituting Eqs.(\ref{eqn:LDM}) and (\ref{eqn:scs}) into Eq.(\ref{eqn:fac}),
we can solve the SDC for $^3P_J^{[1]}$ as
\be
df_{^3P_J^{[1]}}^{NLO}=df^F_{^3P_J^{[1]}}-\frac{\alpha_s}{3\pi m_{c}^{2}}\frac{N_{c}^{2}-1}{N_{c}^{2}}u_{\epsilon}d f_{^3S_1^{[8]}}^{LO},
\label{eqn:soft}
\ee
where
\be
u_\epsilon=u_{\epsilon}^{s}-u_{\epsilon}^{c} ,
\ee
the expressions of which for the $\mu_\Lambda$-cutoff and $\overline{MS}$ renormalization scheme are
\be
u_\epsilon|_{\mu_\Lambda}=\frac{E}{p}ln(\frac{E+p}{E-p})+ln(\frac{\mu_{\Lambda}^{2}}{s\delta_{s}^{2}})-2+2ln(2)
\ee
and
\be
u_\epsilon|_{\overline{MS}}=\frac{E}{p}ln(\frac{E+p}{E-p})+ln(\frac{\mu_{\Lambda}^{2}}{s\delta_{s}^{2}}) ,
\ee
respectively.

Both of the terms on the right-hand side of Eq.(\ref{eqn:soft}) are finite.
Now, all the short-distance coefficients are IR divergence free;then
the components for calculating the cross section for $\chi_c$ hadroproduction are well defined.

Substituting Eq.(\ref{eqn:soft}) into Eq.(\ref{eqn:s}), we obtain the complete expression of the cross section for $\chi_c$ production:
\be
d\sigma^{NLO}(\chi_{cJ})=d\sigma^F(\chi_{cJ})
-\frac{\alpha_s}{3\pi m^{2}_{c}}\frac{N_{c}^{2}-1}{N_{c}^2}u_{\epsilon}\langle O^{\chi_{cJ}}(^3P_J^{[1]})\rangle df_{^3S_1^{[8]}}^{LO}
+(2J+1)\langle O^{\chi_{c0}}(^3S_1^{[8]})\rangle df_{^3S_1^{[8]}}^{NLO} .
\label{eqn:com}
\ee

\section{numerical calculation and the analysis on $\mu_\Lambda$ dependence}

To calculate $\sigma(^3S_1^{[8]})$ and $\sigma(^3P_J^{[1]})$,
we apply our Feynman diagram calculation package (FDC)~\cite{Wang:2004du} to generate the entire needed FORTRAN source.

Before we present the numerical results, we should comment on the obtaining
the CO LDME.
Focusing on the last two terms in the right-hand side of Eq.(\ref{eqn:com}),
one can notice that, if $\mu_\Lambda$ varies its value,
in order to fit the cross section $d\sigma^{NLO}(\chi_{c})$ to the experimental data,
the LDME in the last term should change accordingly,
which is to say, the dependence on $\mu_\Lambda$ is partly absorbed into the CO LDME.
If we proceed with our calculation to infinite order in $\alpha_s$, the $\mu_\Lambda$ dependence can be totally absorbed into the CO LDME.
Consequently, this scale actually can be any positive value holding the convergence of $\alpha_s$ expansion.
If our results significantly depend on $\mu_\Lambda$, the dropped terms in higher orders must contribute significantly,
and the calculation up to this order does not reach a sufficient accuracy.
Up to NLO, the condition of $\mu_\Lambda$ independence requires
\be
\frac{\alpha_{s}}{3\pi m_{c}^{2}}\frac{N_{c}^{2}-1}{N_{c}^{2}}d f^{LO}_{^3S_1^{[8]}}\propto df^{NLO}_{^3S_1^{[8]}} , \label{eqn:cond}
\ee
as well as that the proportional ratio should be universal for all the processes.
We define a quantity
\be
r=\frac{df^{NLO}_{^3S_1^{[8]}}}{dp_t}/(\frac{\alpha_{s}}{3\pi}\frac{N_{c}^{2}-1}{N_{c}^{2}}\frac{d f^{LO}_{^3S_1^{[8]}}}{dp_t}), \label{eqn:defr}
\ee
to determine whether the $\mu_\Lambda$ dependence is severe.
If $r$ is constrained in a small range throughout the whole $p_t$ region for all the processes,
we know for sure the dependence on $\mu_\Lambda$ can be absorbed into the LDME and vice versa.
In this paper, we provide the values of the CO LDME for different renormalization schemes and $\mu_\Lambda$ choices.

In the numerical calculation, we have the following common choices of parameters:
$|R_p'(0)|^{2}=0.075\gev^5$~\cite{Eichten:1995ch} for both LO and NLO calculation, and $m_{c}=1.5\gev$.
The soft cutoff $\delta_s$ independence is checked in the calculation and $\delta_s=0.001$ is used.
Since the energy scale of most of the phase space region exceeds b-quark mass, $\Lambda_{QCD}|_{nf=5}=0.226\gev$ is used.
We employ CTEQ6M~\cite{Pumplin:2002vw} as the parton distribution function (PDF) and two-loop $\alpha_s$ running for up-to-NLO calculation,
and CTEQ6L1~\cite{Pumplin:2002vw} and one-loop $\alpha_s$ running for LO.
The renormalization and factorization scales are chosen as $\mu_R=\mu_f=m_\perp\equiv\sqrt{4m_c^2+p_t^2}$.

In our fit, we exclude the $p_t<7\gev$ data points.
This is because, for one thing,
relativistic correction contributes a part proportional to the QCD LO SDC, when $p_t$ is larger than about 7GeV~\cite{Xu:2012am};
this part can be complemented by modifying the values of the LDMEs,
while below 7GeV, this is not true.
For another thing, the $log(x)$ terms might ruin the perturbative expansion in this region~\cite{Kang:2013hta, Ma:2014mri}.

To extract the CO LDME, we employ all the existing data on $\chi_c$ hadroproduction except for those in Ref.~\cite{Abe:1997yz},
which measured the fraction of the $J/\psi$ hadroproduction cross section through the $\chi_c$ feeddown to the prompt one.
Reference~\cite{Abe:1997jz} provided the prompt $J/\psi$ hadroproduction cross section with the same center-of-mass energy and rapidity range.
However, the $p_t$'s of the two sets of data do not coincide.Thus,
we cannot extract the exact central values and error bars of the $\chi_c$ hadroproduction cross sections and
consequently we cannot use these data directly.
For this reason, we give up using them in our fit.
There are six sets of data involved in our analysis.
All of them are listed in Table~\ref{table:data}.
Except for the data mentioned above, we also noticed another set of data (denoted as E3A), which was published in Ref.~\cite{LHCb:2012ac}.
Since the data in E3 are the updated version of those in E3A, we do not use E3A for fit;however, we plot them in the figures for reference.

\begin{table}[htbp]
\begin{center}
\begin{tabular}{|c|c|c|c|c|c|c|}
\hline
Abbreviation&E1&E2&E3&E4&E5&E6\\
\hline
Center-of-mass energy(TeV)&1.96 &7 &7 &7 &7 &7 \\
\hline
Rapidity range&$|y|<1.0$&$2.0<y<4.5$&$2.0<y<4.5$&$|y|<1.0$&$|y|<0.75$&$|y|<0.75$\\
\hline
Collaboration&CDF~\cite{Abulencia:2007bra}&LHCb~\cite{LHCb:2012af}&LHCb~\cite{Aaij:2013dja}&CMS~\cite{Chatrchyan:2012ub}&ATLAS~\cite{ATLAS:2014ala}&ATLAS~\cite{ATLAS:2014ala}\\
\hline
Content&$\sigma(\chi_{c2})/\sigma(\chi_{c1})$&$\chi_c$ cross section&$\sigma(\chi_{c2})/\sigma(\chi_{c1})$&$\sigma(\chi_{c2})/\sigma(\chi_{c1})$&$\chi_{c1}$ cross section&$\chi_{c2}$ cross section\\
\hline
\end{tabular}
\caption{
All the sets of data used in our fit.
}
\label{table:data}
\end{center}
\end{table}

In Refs.~\cite{Abulencia:2007bra, LHCb:2012af, Aaij:2013dja, Chatrchyan:2012ub} (corresponding to E1$-$E4), the values of $p_t$ are given for the $J/\psi$ generated from $\chi_c$ feeddown, in accordance with which
we should do the so-called $p_t$ shift as $p_t^{J/\psi}\approx p_t^{\chi_{cJ}}m_{J/\psi}/m_{\chi_{cJ}}$.
Here we choose~\cite{Agashe:2014kda} $m_{J/\psi}=3.097\gev$, $m_{\chi_{c0}}=3.415\gev$, $m_{\chi_{c1}}=3.510\gev$,and $m_{\chi_{c2}}=3.556\gev$,
which are different from Ref.~\cite{Gong:2012ug}, where $m_{J/\psi}$ and $m_{\chi_c}$ are $3.1\gev$ and $3.5\gev$, respectively.
The branching ratios~\cite{Agashe:2014kda} are $1.27\%$, $33.9\%$, and $19.2\%$ for $\chi_{c0,1,2}$ to $J/\psi$, respectively.

\begin{figure}
\center{
\includegraphics*[scale=0.5]{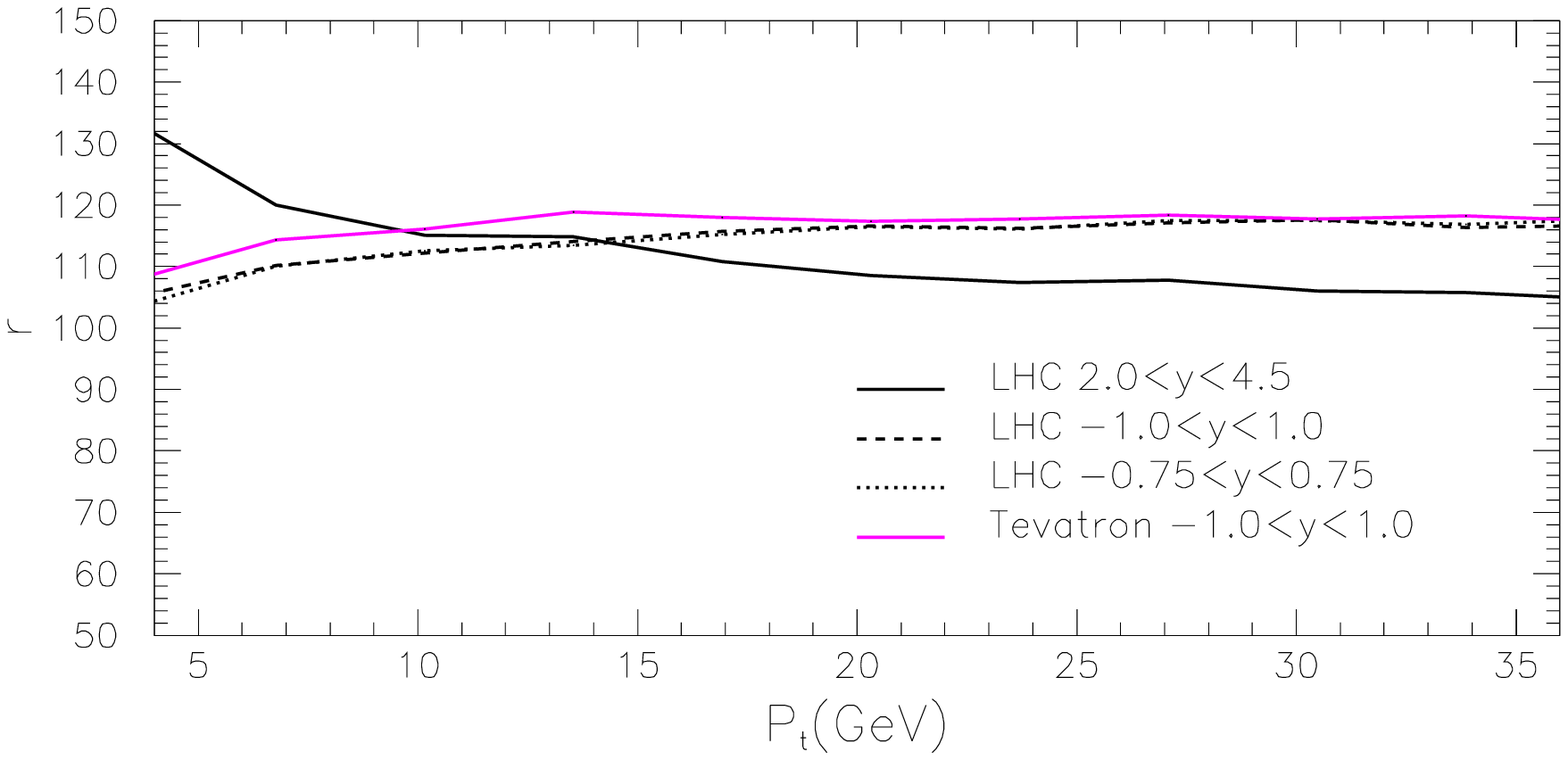}
\caption{\label{fig:ratio} The value of $r$ defined in Eq.(\ref{eqn:defr}) as a function of $p_t^{\chi_c}$.}
}
\end{figure}

Before we carry out the fit, we shall first investigate whether $r$ defined in Eq.(\ref{eqn:defr})
is universally a constant to hold the $\mu_\Lambda$ independence.
E2 and E3 are in the same experimental condition, as are E5 and E6; accordingly, there are four conditions to present.
For these experimental conditions, the values of $r$ are presented in Fig.\ref{fig:ratio}.
We can see that, except for E2 (as well as E3), for all three conditions, $r$ is almost a constant,
ranging from about 104 to 118, as $p_t$ varies from $4\gev$ to $36\gev$.
As we expected in the introductory section that NLO results should provide a sufficiently precise prediction,
the $\mu_\Lambda$ dependence cannot be severe.
For E2, the situation is a little worse ($r$ ranges from 132 to 105), yet,
not so bad to ruin the results.
We can now expect that, for E1 and E4$-$E6, the theoretical prediction should be in good agreement with the experiment,
and if we carry out the fit by employing the four sets of data individually, we should obtain almost the same results of the values of the LDME.
For E2 and E3, the theoretical prediction might agree with the experiment qualitatively.

$r$ is not always a constant up to NLO precision.
As an example, for $h_c$ hadroproduction, $r$ varies significantly for different experimental conditions and phase space regions.
Interested readers can refer to Ref.~\cite{Wang:2014vsa}, in which the detailed results for $h_c$ hadroproduction are presented.

\section{Numerical results and comparison to the experimental data}

We present the values of the CO LDME extracted from the fit of each set of the experimental data at both LO and NLO,
and see if they correspond with one another.
In the rest of this paper, we abbreviate $\langle O^{\chi_{c0}}(^3S_1^{[8]})\rangle$ as ${\cal O}\times 10^{-3}\gev^3$.

For LO calculation,
\bea
&&{\cal O}_{E1}^{LO}=0.24\pm 0.13,~~~~~~{\cal O}_{E2}^{LO}=1.26\pm 0.03, \NO \\
&&{\cal O}_{E3}^{LO}=0.19\pm 0.06,~~~~~~{\cal O}_{E4}^{LO}=0.13\pm 0.05, \label{eqn:LDMELO} \\
&&{\cal O}_{E5}^{LO}=1.22\pm 0.07,~~~~~~{\cal O}_{E6}^{LO}=0.67\pm 0.07, \NO
\eea
and the $\chi^2/d.o.f.$ are 6.4, 0.0078, 1.1, 0.69, 0.11, and 0.43, respectively.

For NLO calculation, as $\mu_\Lambda=m_c$ in $\mu_\Lambda$-cutoff renormalization scheme,
\bea
&&{\cal O}_{E1}^{NLO}=1.97\pm 0.17,~~~~~~{\cal O}_{E2}^{NLO}=2.34\pm 0.06, \NO \\
&&{\cal O}_{E3}^{NLO}=2.28\pm 0.06,~~~~~~{\cal O}_{E4}^{NLO}=2.00\pm 0.07, \label{eqn:LDMENLO} \\
&&{\cal O}_{E5}^{NLO}=2.03\pm 0.05,~~~~~~{\cal O}_{E6}^{NLO}=2.04\pm 0.06, \NO
\eea
and the $\chi^2/d.o.f.$ are 2.8, 0.034, 0.18, 0.17, 0.068, and 0.35, respectively.

To begin with, we can see that, for all the conditions except for E2, the $\chi^2$ for NLO is smaller than that for LO.
Moreover, for E1 and E4$-$E6, the obtained values of the CO LDME for NLO are almost the same (${\cal O}$ ranges from 1.97 to 2.04).
As we analyzed in the previous section, we do not expect theoretical prediction for E2 and E3 to agree with the experiment in high precision;
however, even for E2 and E3, the obtained values of the CO LDME are very close to those for E1 and E4$-$E6.
By contrast,for the values of ${\cal O}$ obtained for the LO range from 0.13 to 1.26,
the largest is about ten times the smallest,
and there is no common value for any group of the sets of data; the distribution of the values is dispersive.
We can conclude that, no universal value exists for LO LDME,
since up to LO, the precision is not sufficient to describe all the experiments.
We can also see that the LDME given in Ref.~\cite{Cho:1995ce} is too large.
One might make wrong conclusions if using that value.
The LDME given in Ref.~\cite{Sharma:2012dy} is much larger than the upper bound of the series of the values presented above.
The large value of the LDME might lead to overestimation of the absorption effect (as well as other nuclear matter effects).

Now we carry out a global fit, using all the experimental data in E1$-$E6 for LO,
and E1 and E4$-$E6 for NLO, and obtain
\be
{\cal O}^{LO}=0.31\pm 0.09,~~~~~~{\cal O}_{\mu_\Lambda}^{NLO}=2.01\pm 0.04. \label{eqn:LDMEdefault}
\ee
The $\chi^2/d.o.f.$ are 2.4 and 0.47 for LO and NLO, respectively.
The consideration is that we trust the precision of the NLO results for E1 and E4$-$E6.
However, for E2 and E3, the situation is not clear.
Thus, we fit E1 and E4$-$E6 to obtain NLO CO LDME as a default value to present our results,
and we employ the LDME to see whether it can explain the experiment E2 and E3.

\begin{figure}
\center{
\includegraphics*[scale=0.5]{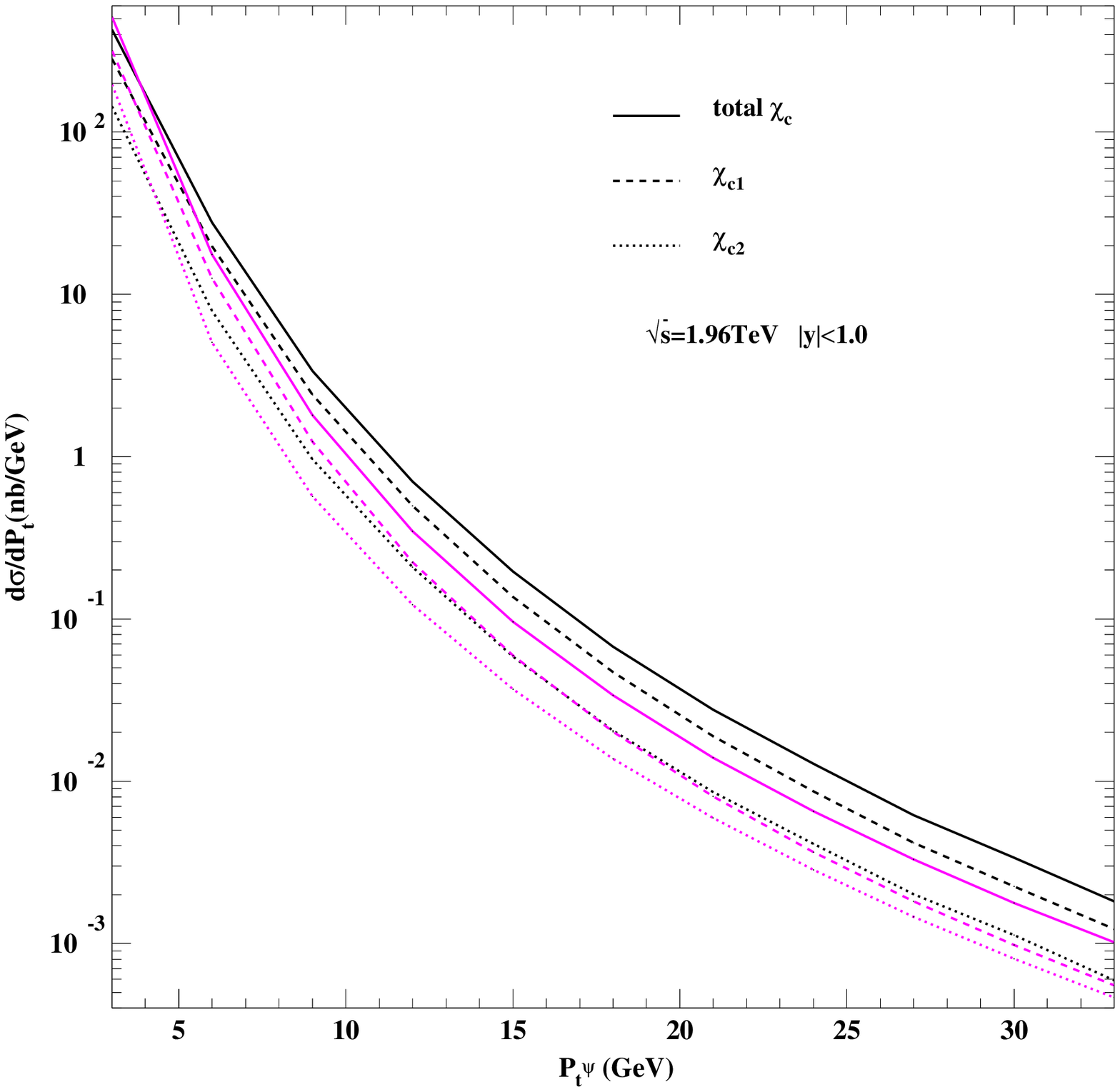}
\includegraphics*[scale=0.5]{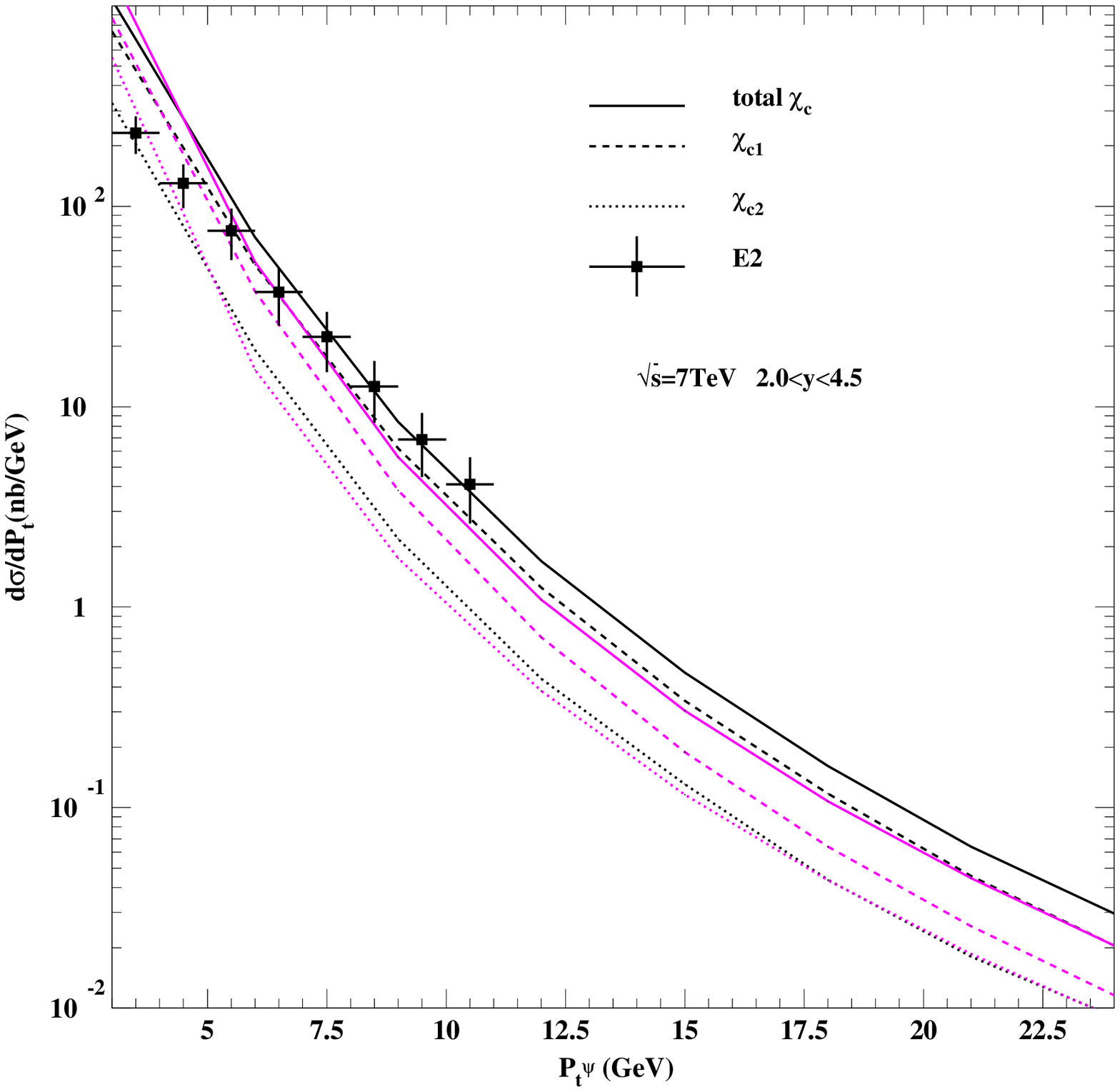}\\
\includegraphics*[scale=0.5]{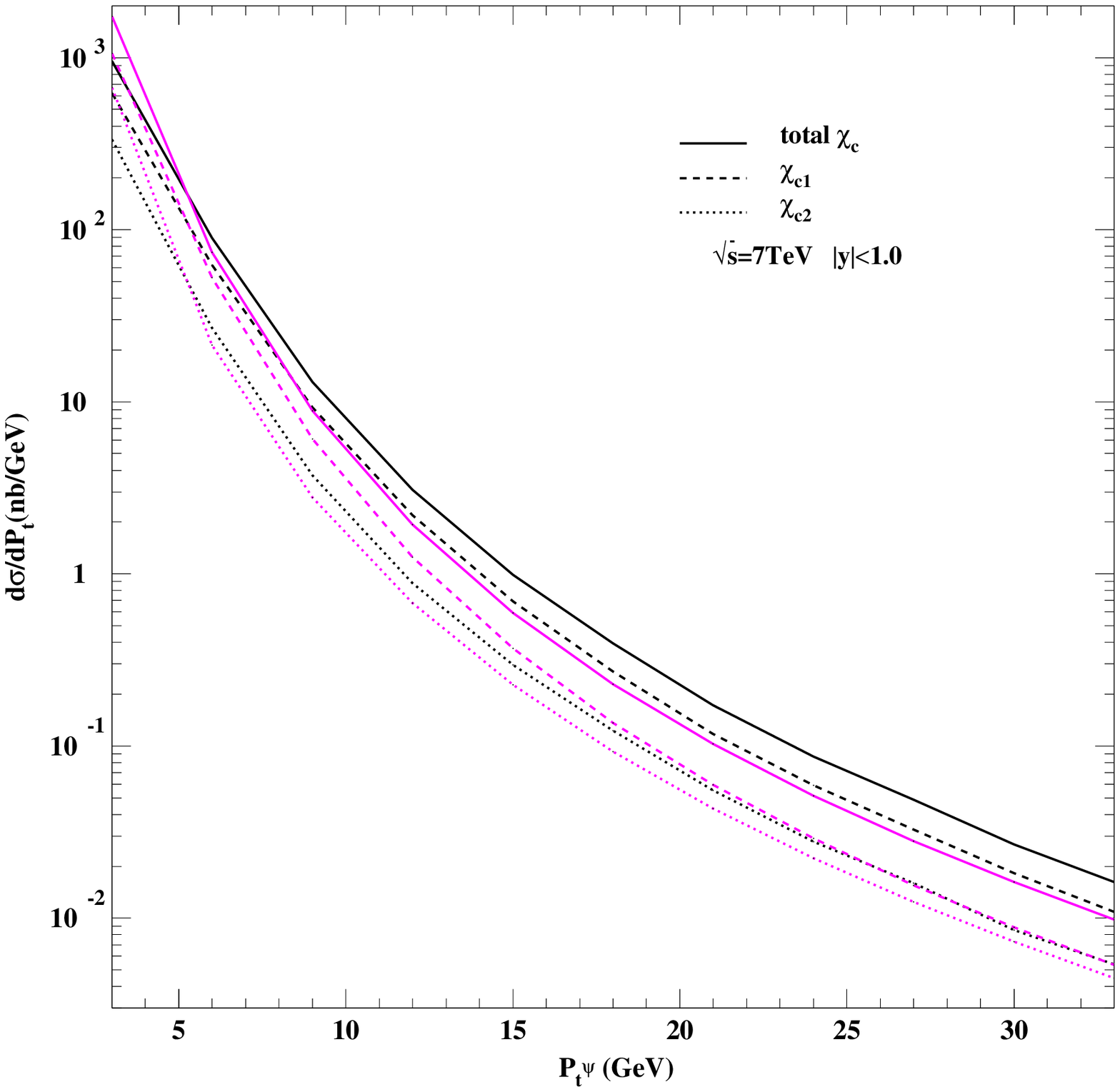}
\includegraphics*[scale=0.5]{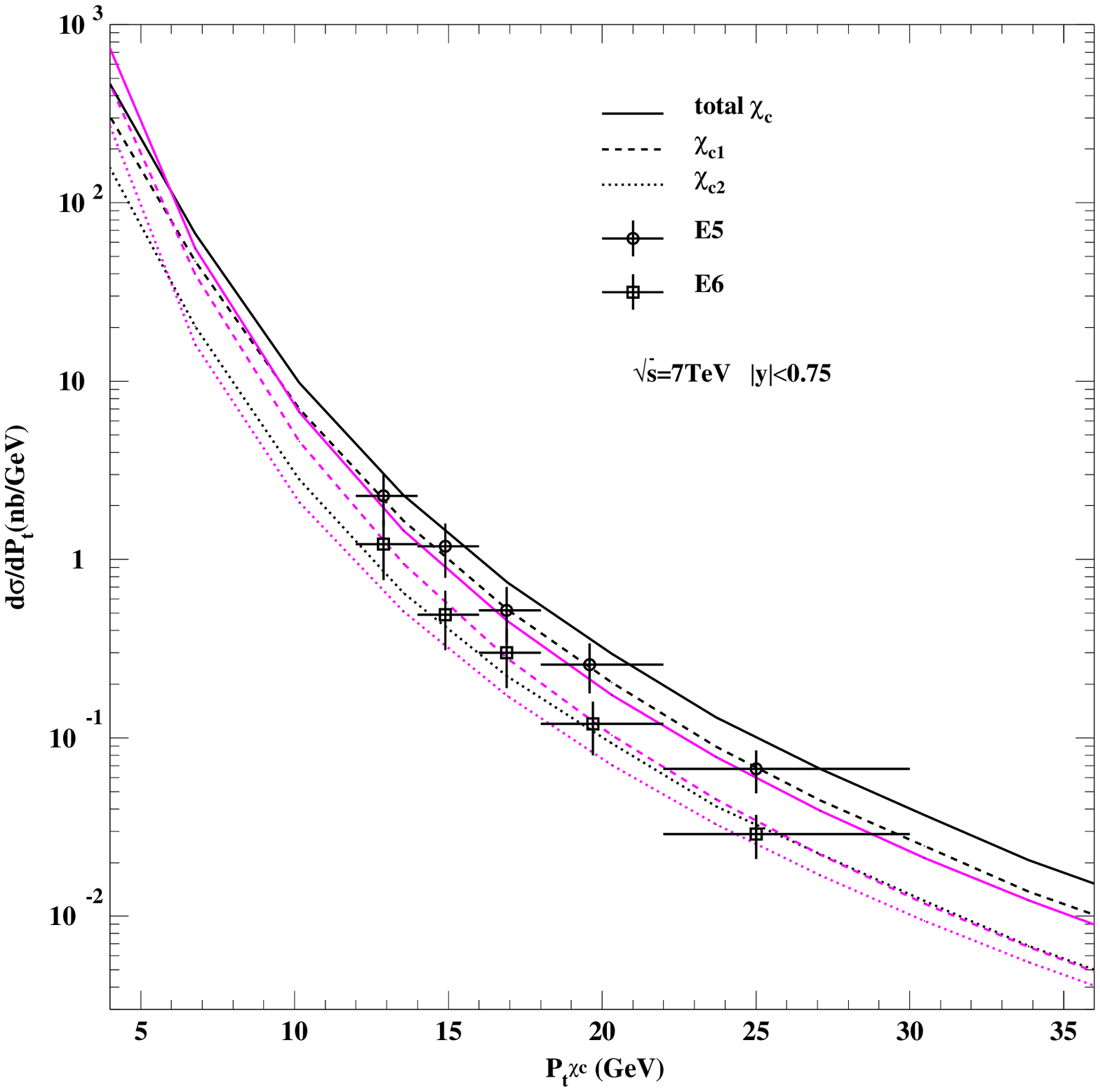}
\caption{\label{fig:pt} The $p_t$ distribution of $\chi_c$ production at the Tevatron and LHC.
The blue and black curves are for LO and NLO, respectively.
The experimental data are taken from Refs.~\cite{LHCb:2012af, ATLAS:2014ala}.}
}
\end{figure}

\begin{figure}
\center{
\includegraphics*[scale=0.5]{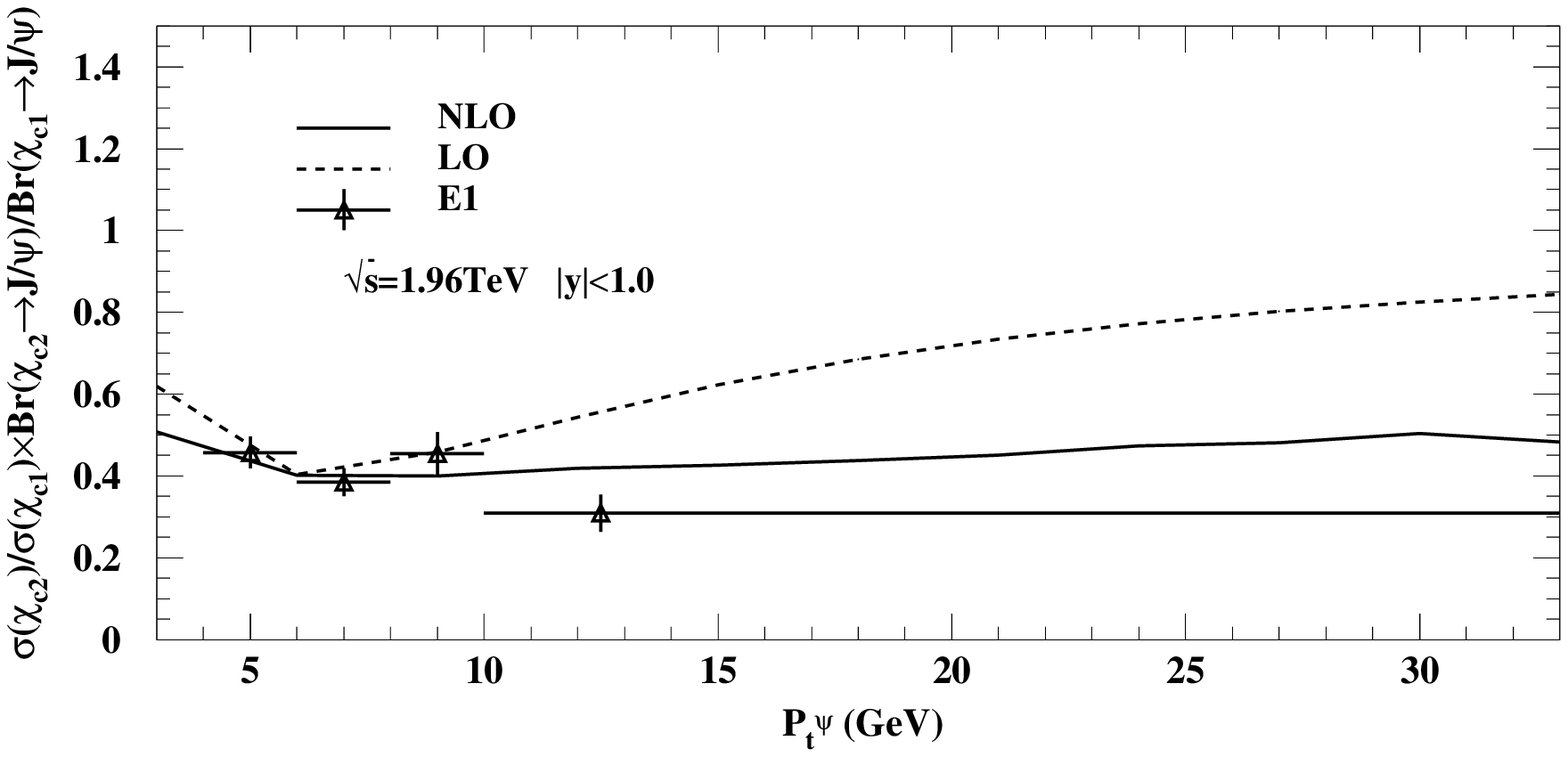}
\includegraphics*[scale=0.5]{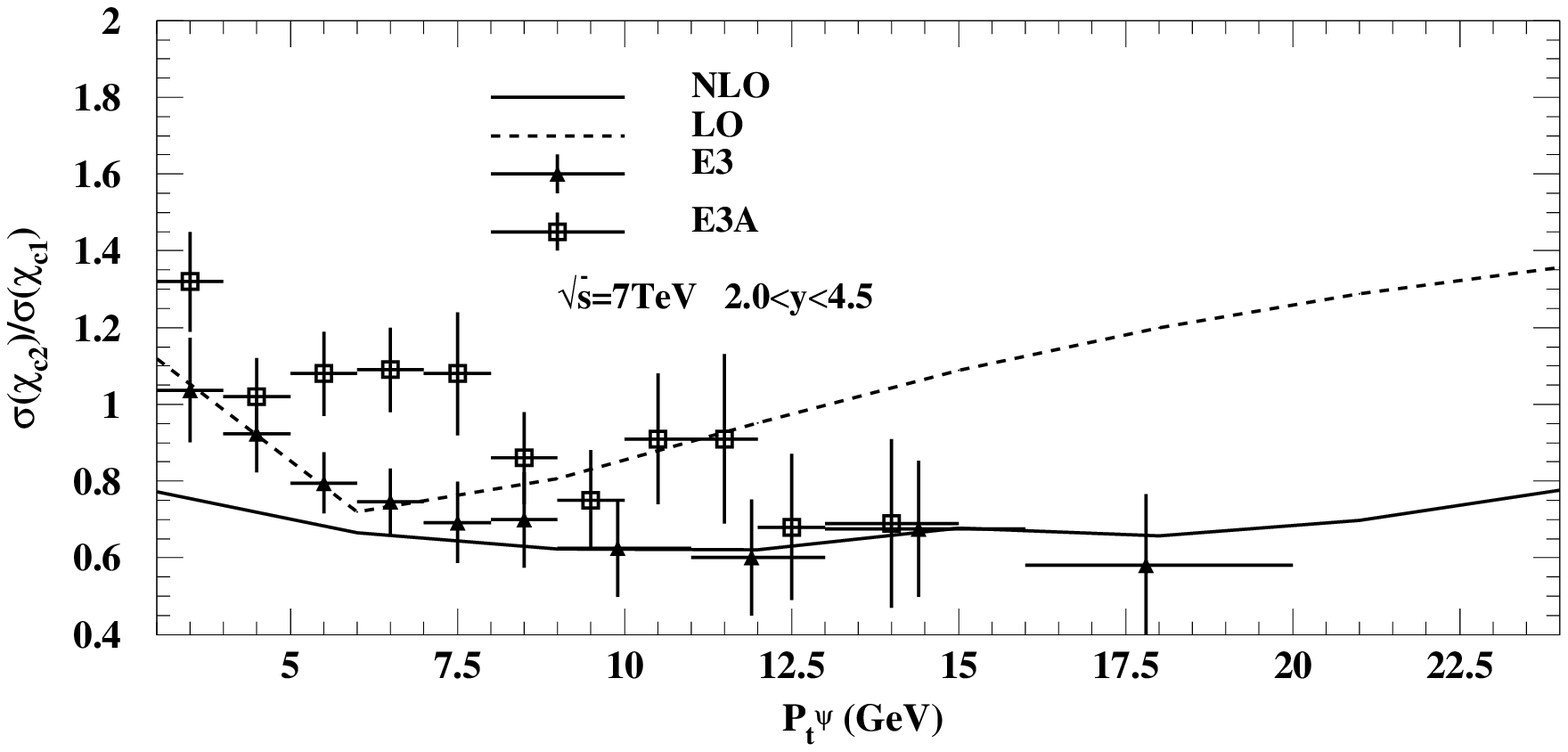}\\
\includegraphics*[scale=0.5]{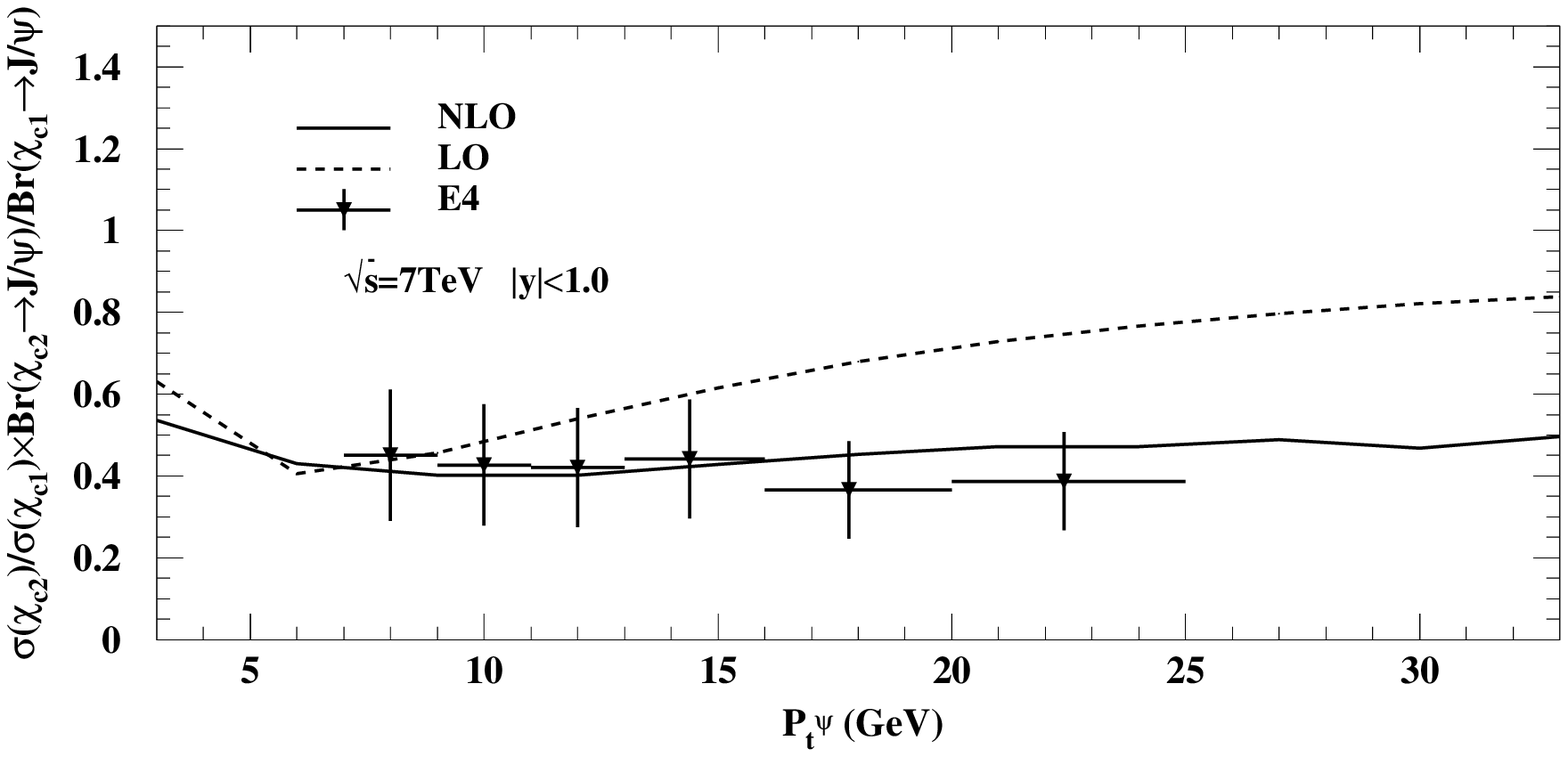}
\includegraphics*[scale=0.5]{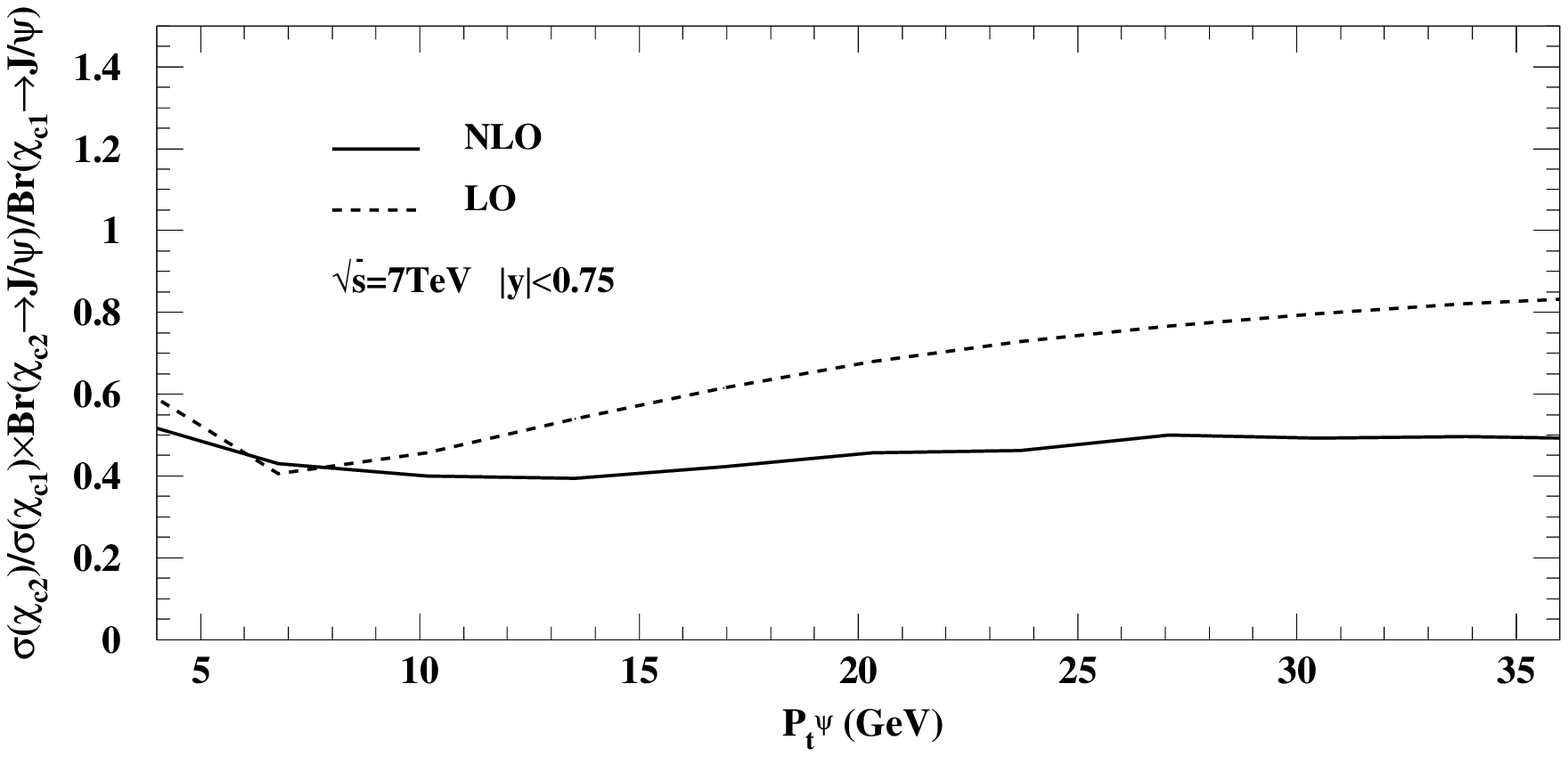}
\caption{\label{fig:r} The ratio $\sigma(\chi_{c2})/\sigma(\chi_{c1})$ as a function of $p_t$ at the Tevatron and LHC.
The blue and black curves are for LO and NLO, respectively.
The experimental data are taken from Refs.~\cite{Abulencia:2007bra, Aaij:2013dja, LHCb:2012ac, Chatrchyan:2012ub}.}
}
\end{figure}

Theoretical predictions for the $\chi_c$ production cross section and the ratio $\sigma(\chi_{c2})/\sigma(\chi_{c1})$ are listed in
Figs.\ref{fig:pt} and \ref{fig:r} , respectively.
References~\cite{Abulencia:2007bra, LHCb:2012af, Aaij:2013dja, Chatrchyan:2012ub} only provide results for $J/\psi$ $p_t$,
while Ref.~\cite{ATLAS:2014ala} provides results for both $J/\psi$ and $\chi_c$ $p_t$.
Since our calculations are carried out at $\chi_c$ $p_t$,
for this reason, the distributions for E5 and E6 illustrated in Fig.\ref{fig:pt} are with respect to $\chi_c$ $p_t$.
Since the uncertainty of the LDME at NLO is small, we do not draw the band rising from this uncertainty.
However, the uncertainty of the LDME for LO is quite large, but we do not bother with this matter here.
We will provide a more reasonable band for LO uncertainty later.

We can see from Figs.\ref{fig:pt} and \ref{fig:r} that NLO results are in very good agreement with all the experiments,
while LO results cannot agree with most of the experimental data.
As we mentioned above, for E2 and E3, the NLO calculation might not be able to provide sufficiently precise results,
since as displayed in Fig.\ref{fig:ratio} that $\mu_\Lambda$ dependence is severe for this experimental condition.
This fact might arise from the large rapidity (denoted as $y$),
since large $y$ eventually introduces two scales $E_{\chi_c}$ (the energy of $\chi_c$) and $m_\perp$.
When $y=4.5$, $E_{\chi_c}/m_\perp\approx 45$, which might ruin the perturbative expansion.
One might resume these terms to achieve well-converged results.

\begin{figure}
\center{
\includegraphics*[scale=0.5]{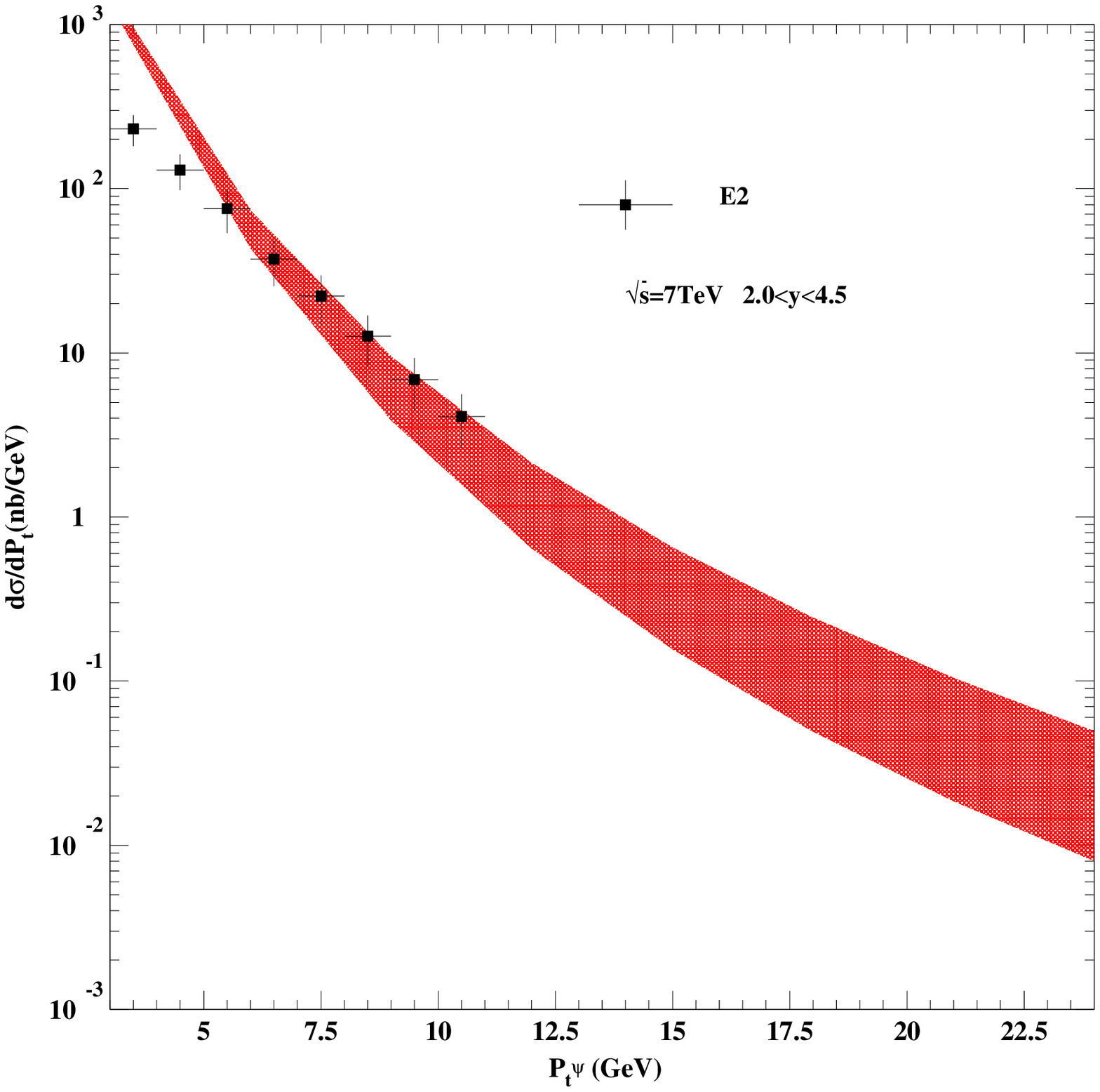}
\includegraphics*[scale=0.5]{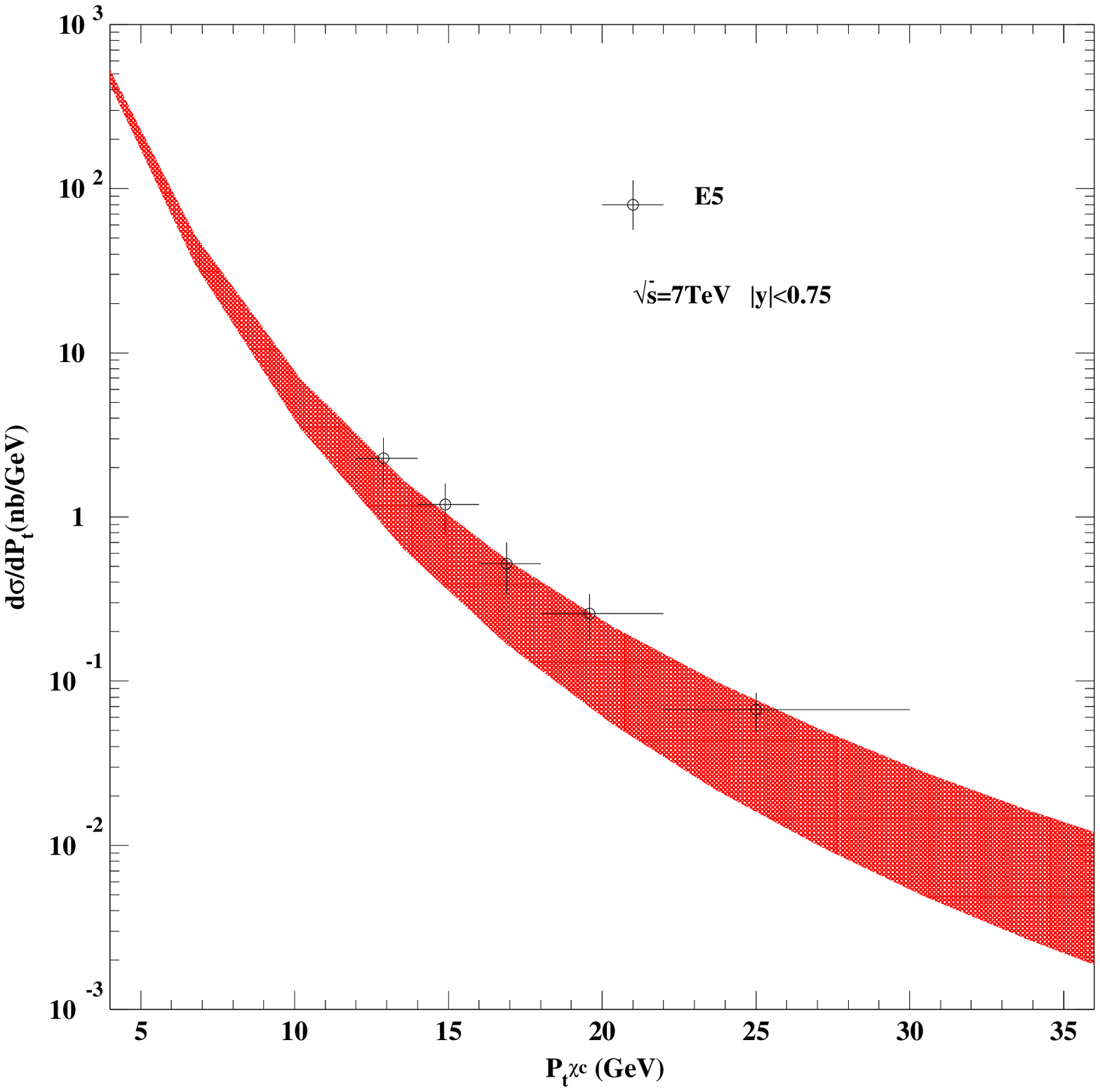}\\
\includegraphics*[scale=0.5]{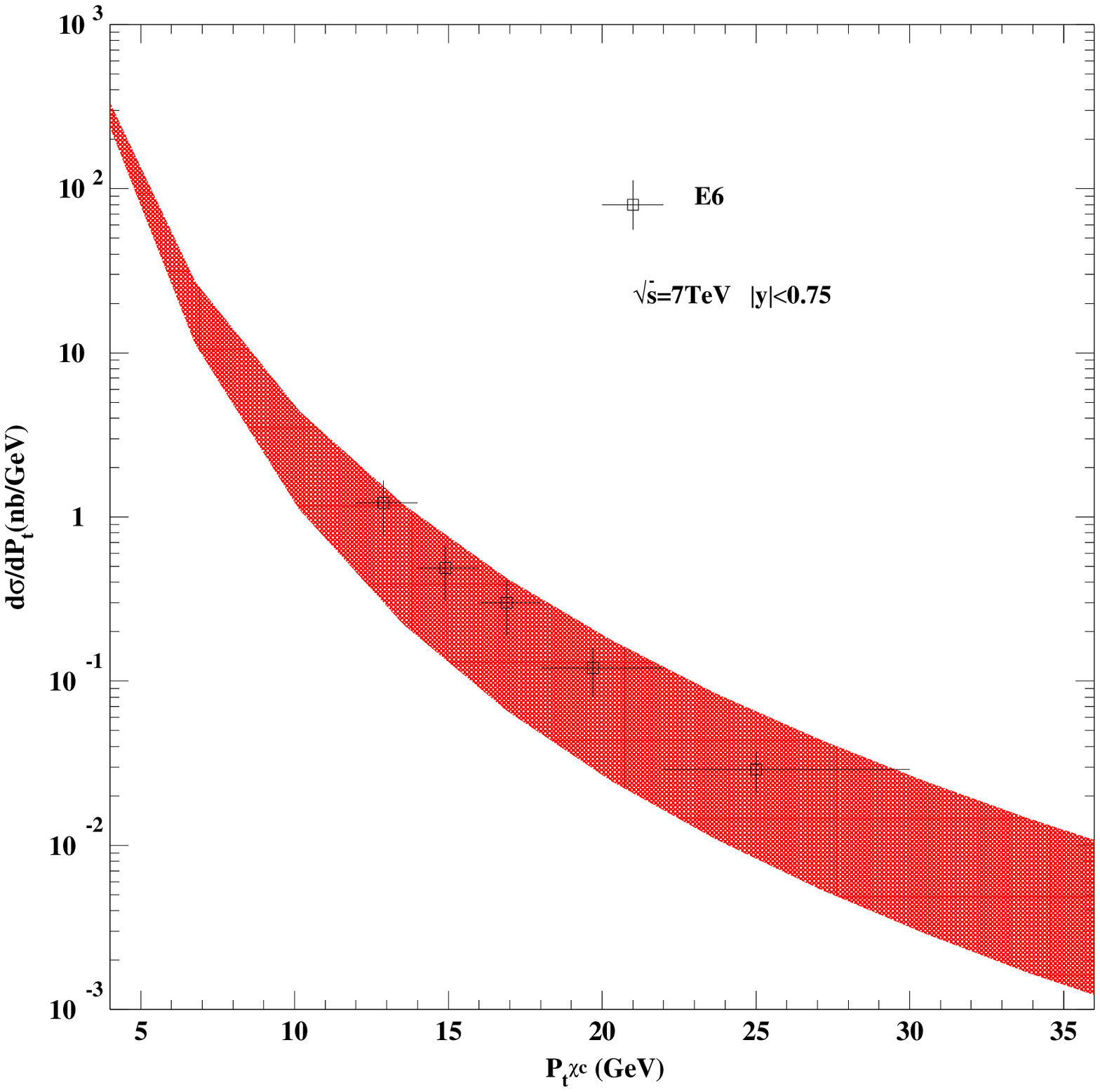}\\
\includegraphics*[scale=0.5]{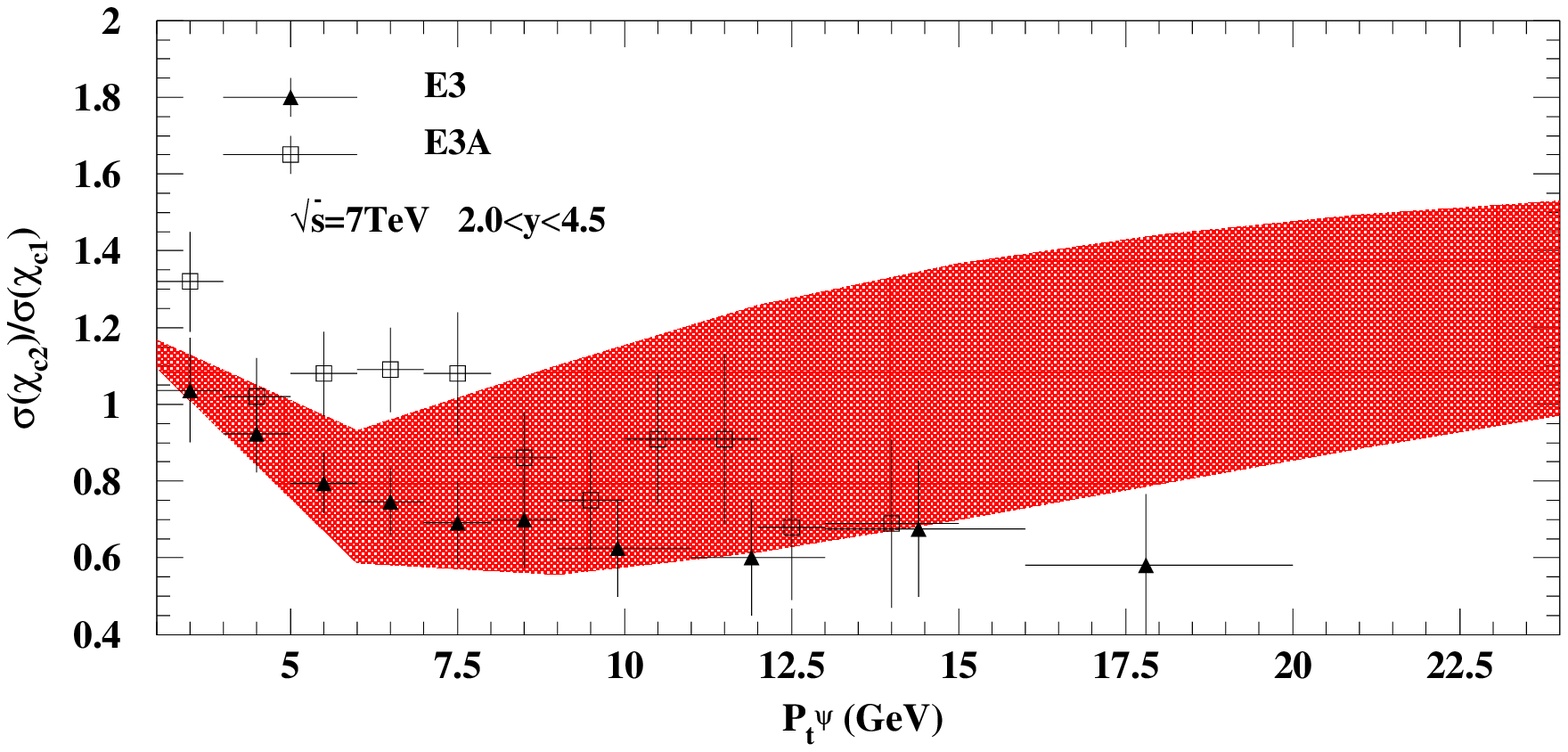}
\includegraphics*[scale=0.5]{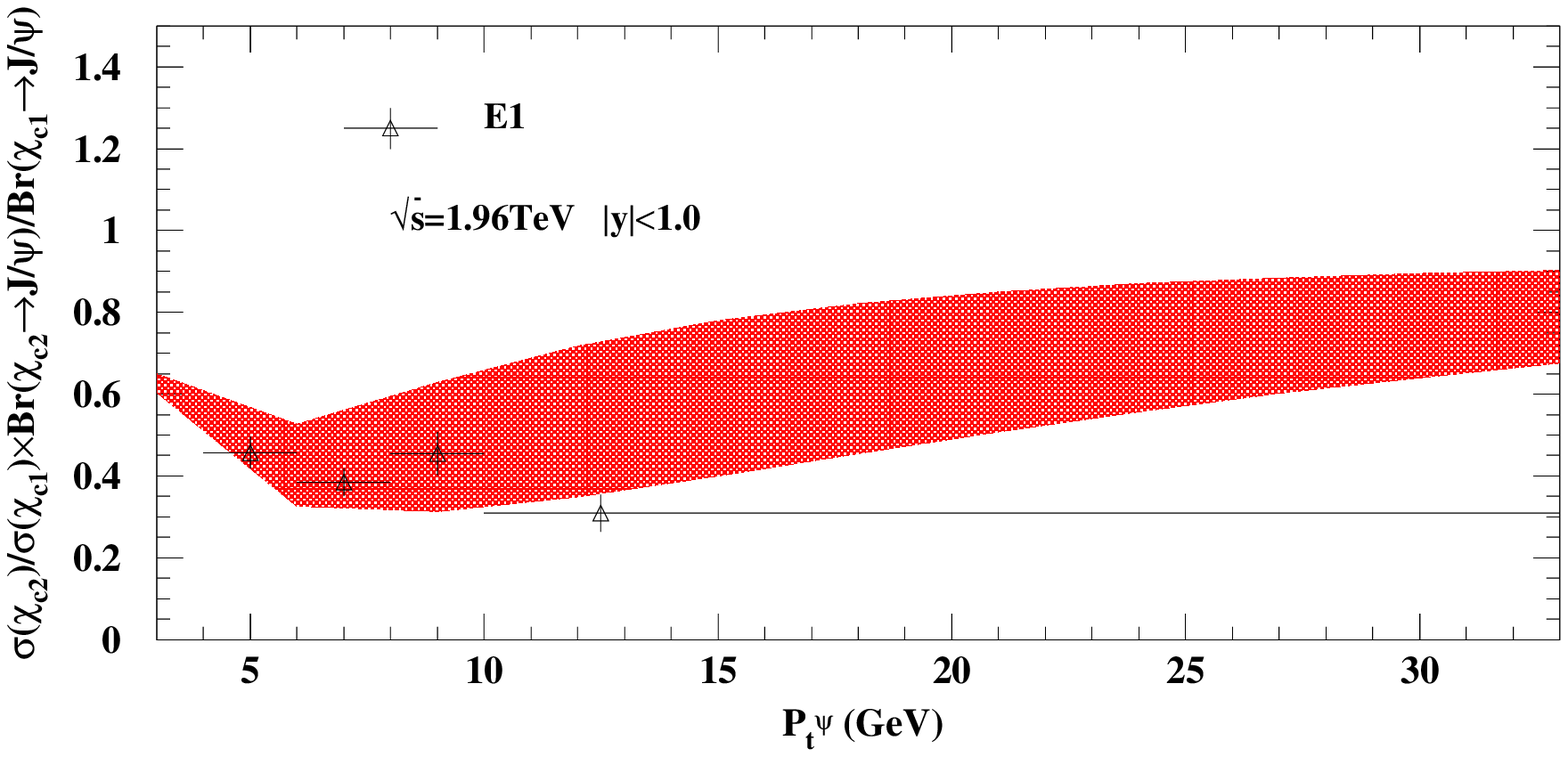}\\
\includegraphics*[scale=0.5]{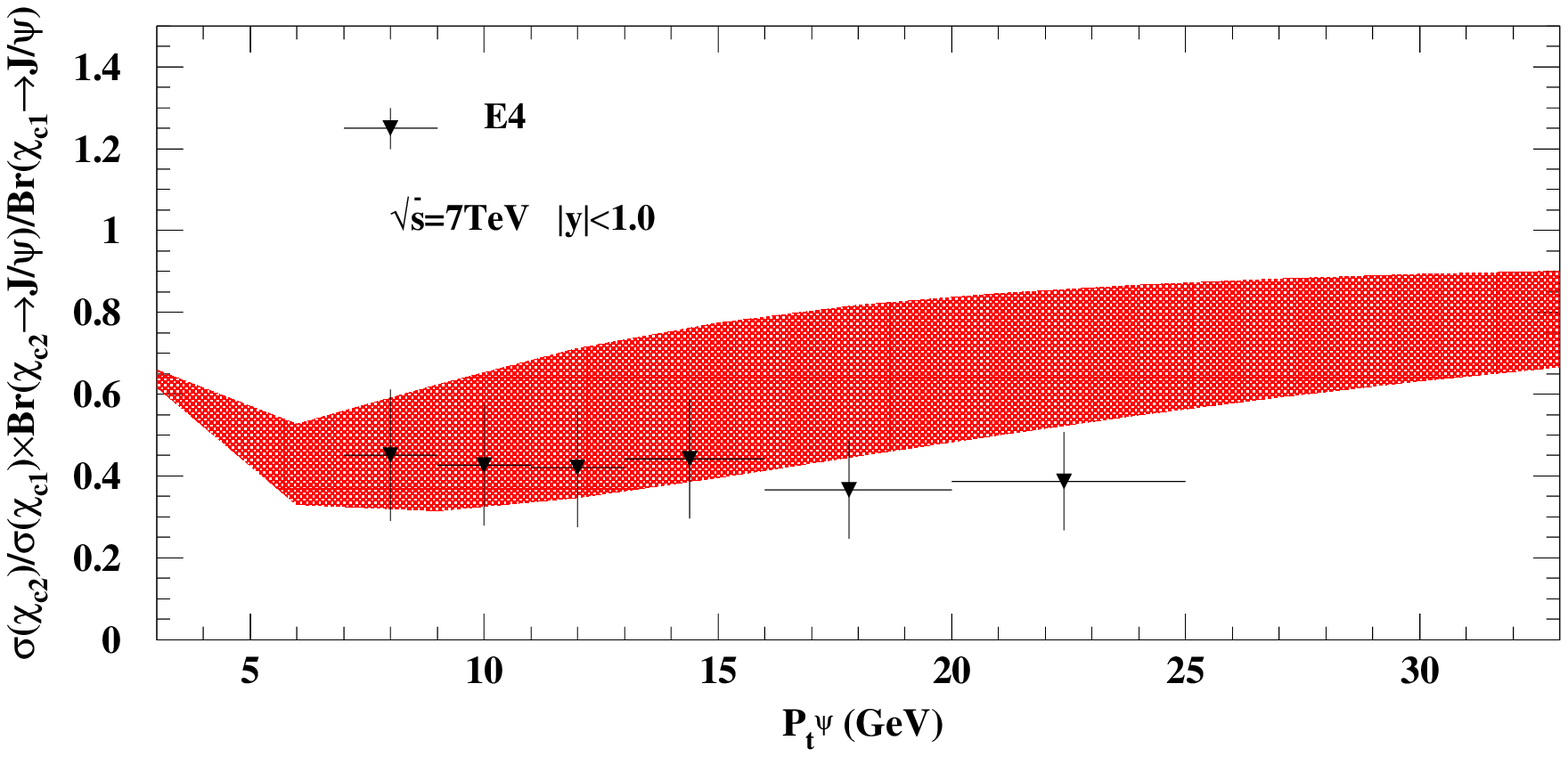}
\caption{\label{fig:lo} The LO results for the $\chi_c$ production cross section and the ratio $\sigma(\chi_{c2})/\sigma(\chi_{c1})$
as a function of $p_t$ at the Tevatron and LHC.
The band corresponds to the LO prediction between the results for ${\cal O}=0.13$ and ${\cal O}=1.88$.
The experimental data are taken from Refs.~\cite{Abulencia:2007bra, LHCb:2012af, Aaij:2013dja, LHCb:2012ac, Chatrchyan:2012ub, ATLAS:2014ala}.}
}
\end{figure}

Since LO results obtained from the default choice of the LDME cannot provide good predictions,
we shall give a range of the LDME to cover all the experimental data.
Here we choose the range between the upper and lower bound of the values given in Eq.(\ref{eqn:LDMELO}): ${\cal O}=0.13\sim 1.26$.
We can see in Fig.\ref{fig:lo} that the large band can cover most of the experimental data,
and the upper bound overestimates the significance of $\chi_c$ feeddown contributions for some of the conditions.
Still, we are not sure whether they are able to explain new experiments; however,
a band presented for the range given above might cover the experimental data in the sense of statistics.
And we know for sure that, a single value of the CO LDME cannot give reasonable predictions at LO.

At the end of this section, we present the values of the CO LDME at NLO for different choices of $\mu_\Lambda$.
For the same reason, we exclude E2 and E3 data.
The LDMEs are listed as follows:
\be
{\cal O}_{m_c}^{NLO}=2.25\pm 0.04,~~~~{\cal O}_{m_c/2}^{NLO}=1.68\pm 0.04,
~~~~{\cal O}_{\Lambda_{QCD}}^{NLO}=0.70\pm 0.04. \label{eqn:LDMEscale}
\ee
Here we have used $\overline{MS}$ renormalization scheme (in the calculation of NLO correction to the CO LDME).
The $\chi^2/d.o.f.$ are 0.48, 0.46, and 0.42, respectively.

Fitting experimental data at different $\mu_\Lambda$'s is actually an alternative procedure of solving the LDME running equation,
which can be obtained from the renormalization of the LDME as
\be
\mu_\Lambda\frac{\partial}{\partial\mu_\Lambda}\langle O^{\chi_{cJ}}(^3S_1^{[8]})\rangle=
\frac{2\alpha_s}{3\pi m_c^2}\frac{N_c^2-1}{N_c^2}\langle O^{\chi_{cJ}}(^3P_J^{[1]})\rangle \label{eqn:running}
\ee
As Fig.\ref{fig:ratio} shows, $r$ is almost a constant for the experimental conditions in the fit,so
we can expect that the LDMEs listed above would be consistent with Eq.(\ref{eqn:running}).
From Fig.\ref{fig:ratio} and Eq.(\ref{eqn:defr}), the typical value of $\alpha_s$ is about 0.09 (at large $p_t$).
Employing this value, we find that the LDMEs in Eq.(\ref{eqn:LDMEscale}) satisfy Eq.(\ref{eqn:running}).

\begin{figure}
\center{
\includegraphics*[scale=0.5]{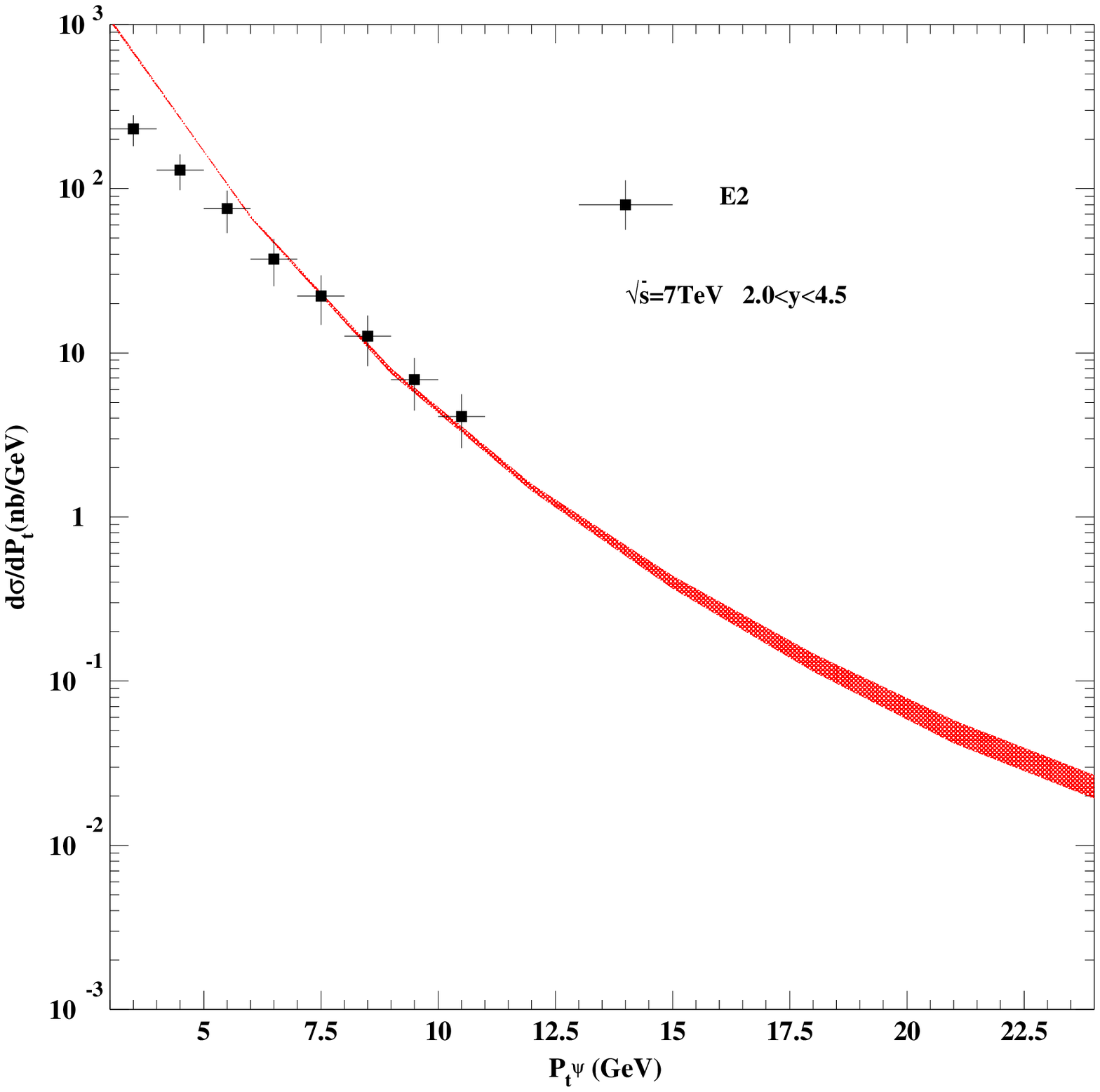}
\includegraphics*[scale=0.5]{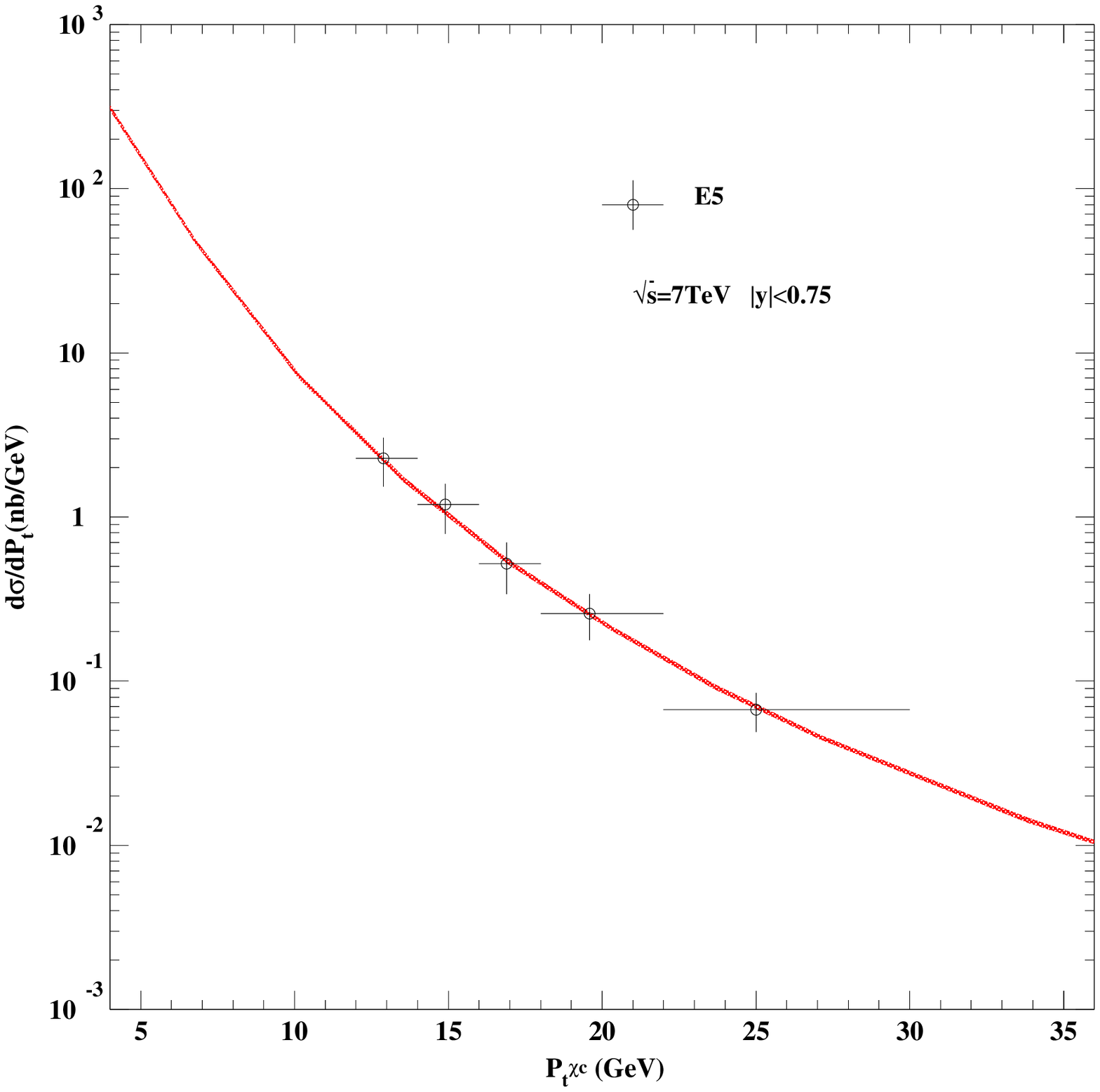}\\
\includegraphics*[scale=0.5]{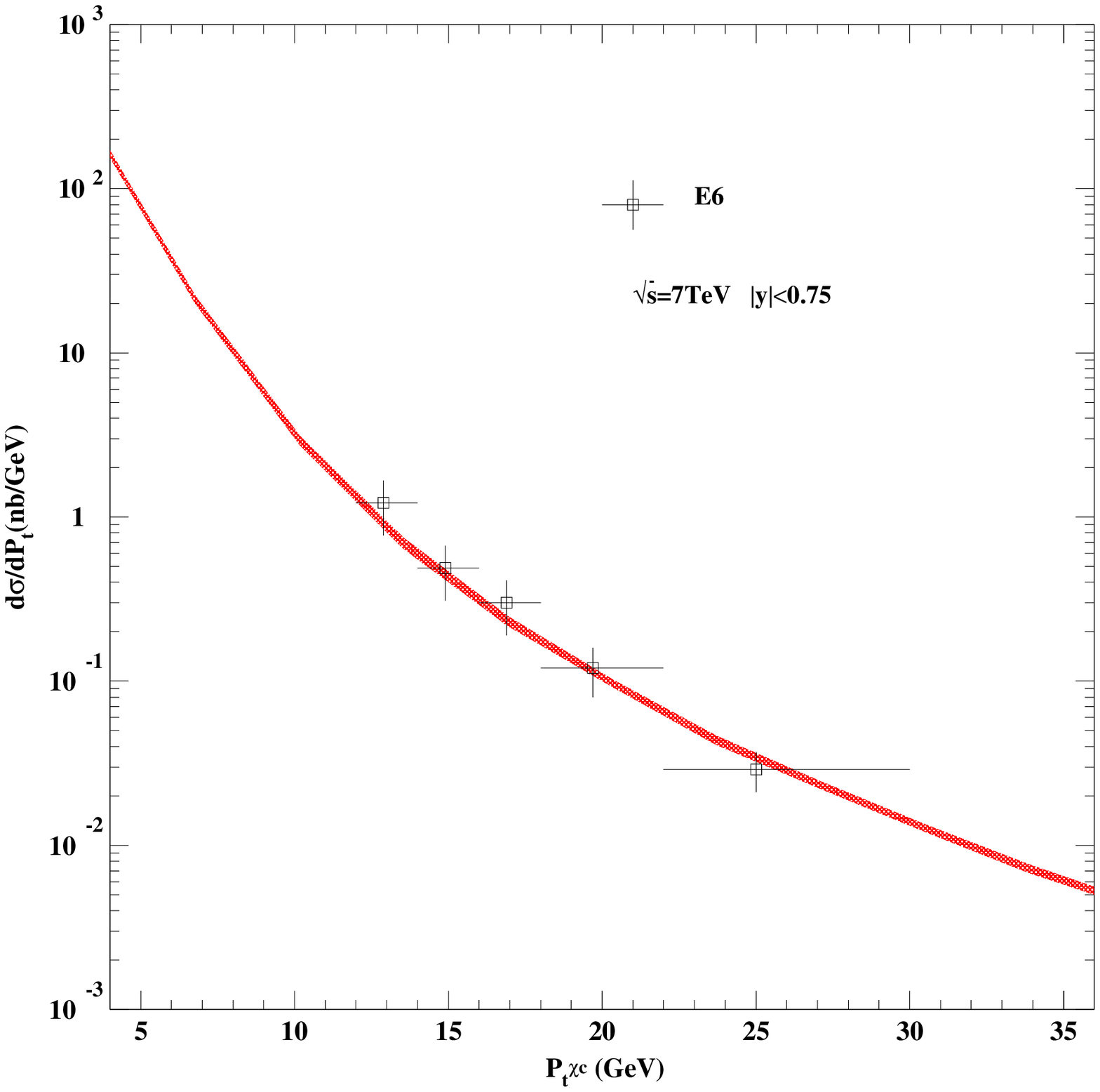}\\
\includegraphics*[scale=0.5]{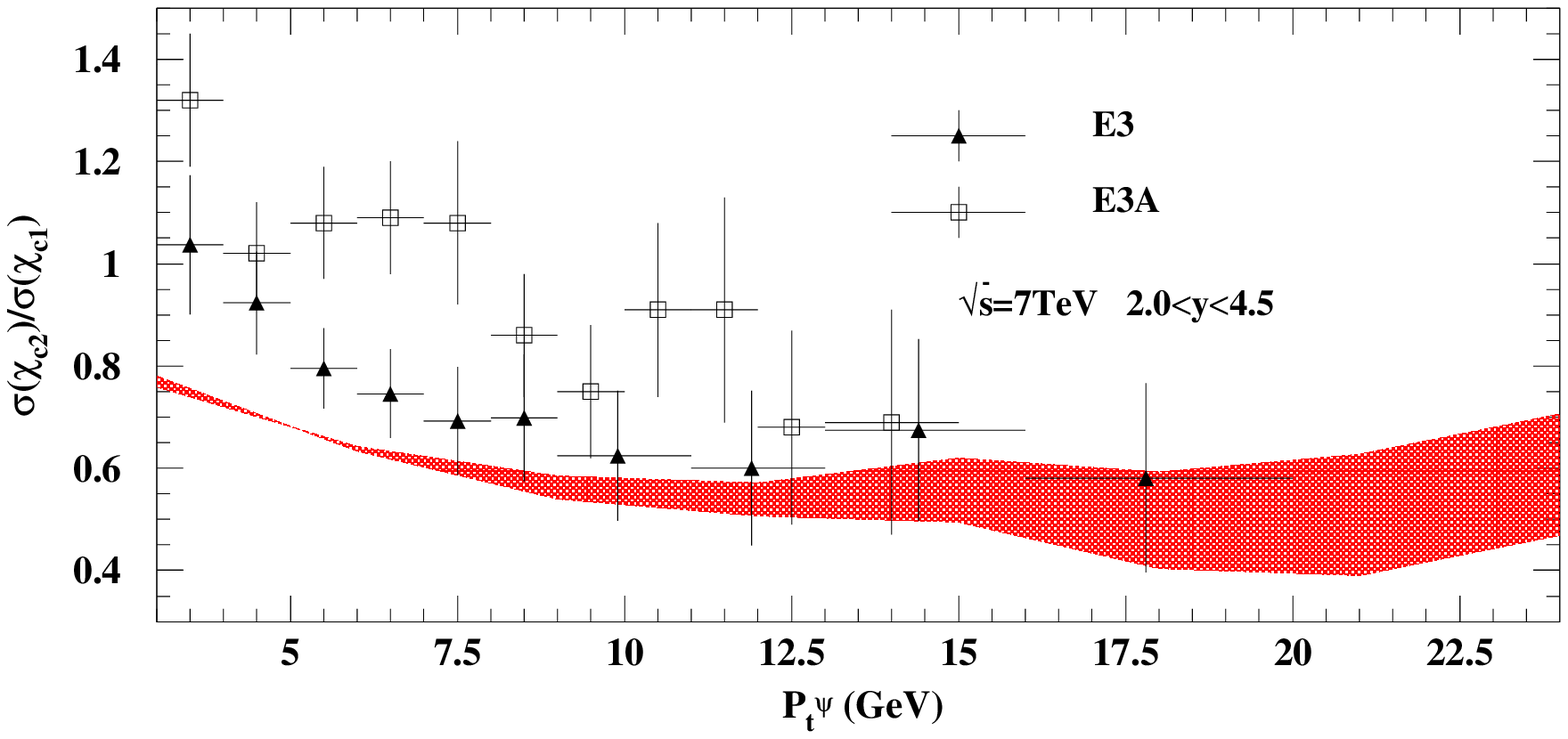}
\includegraphics*[scale=0.5]{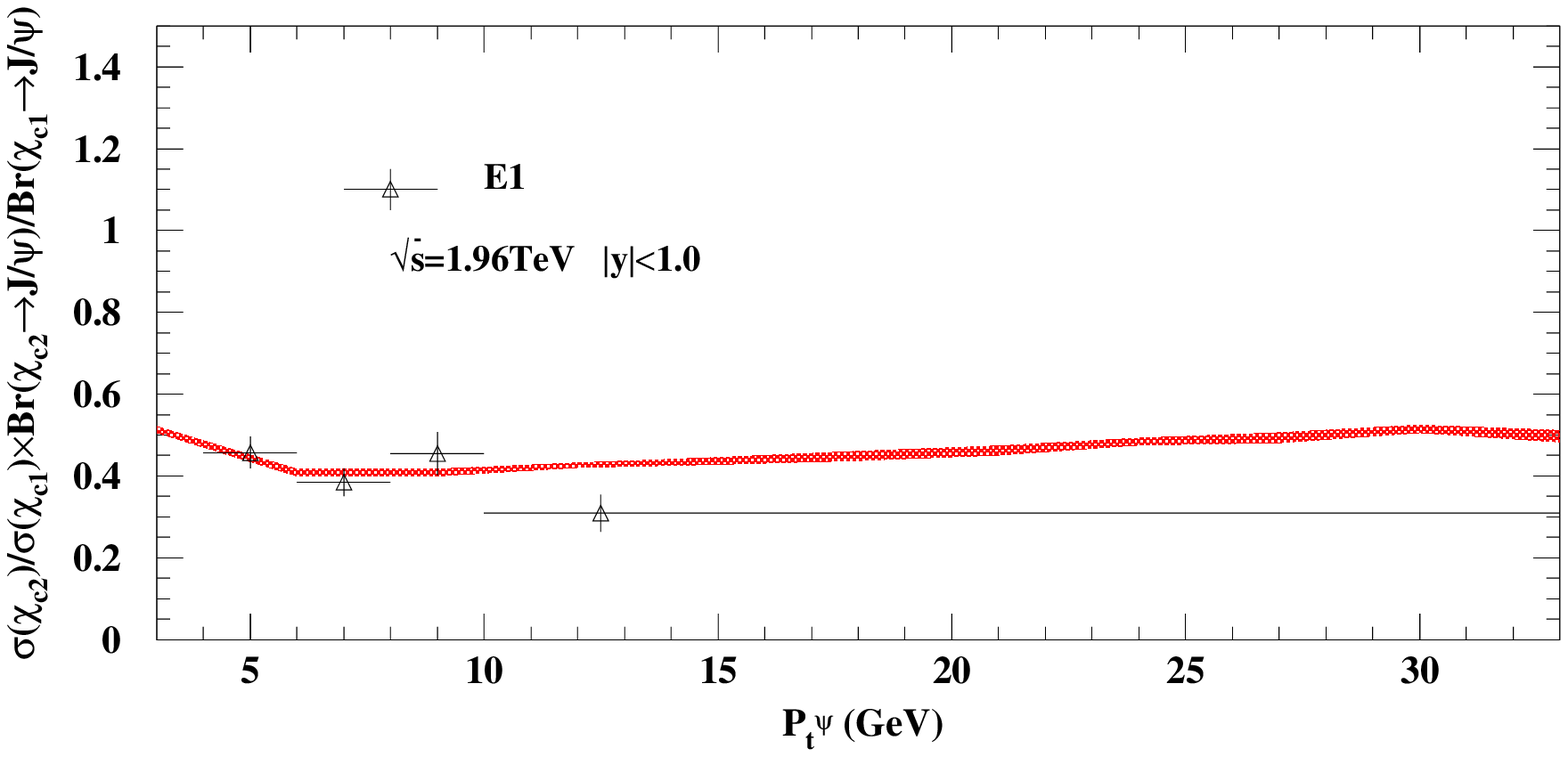}\\
\includegraphics*[scale=0.5]{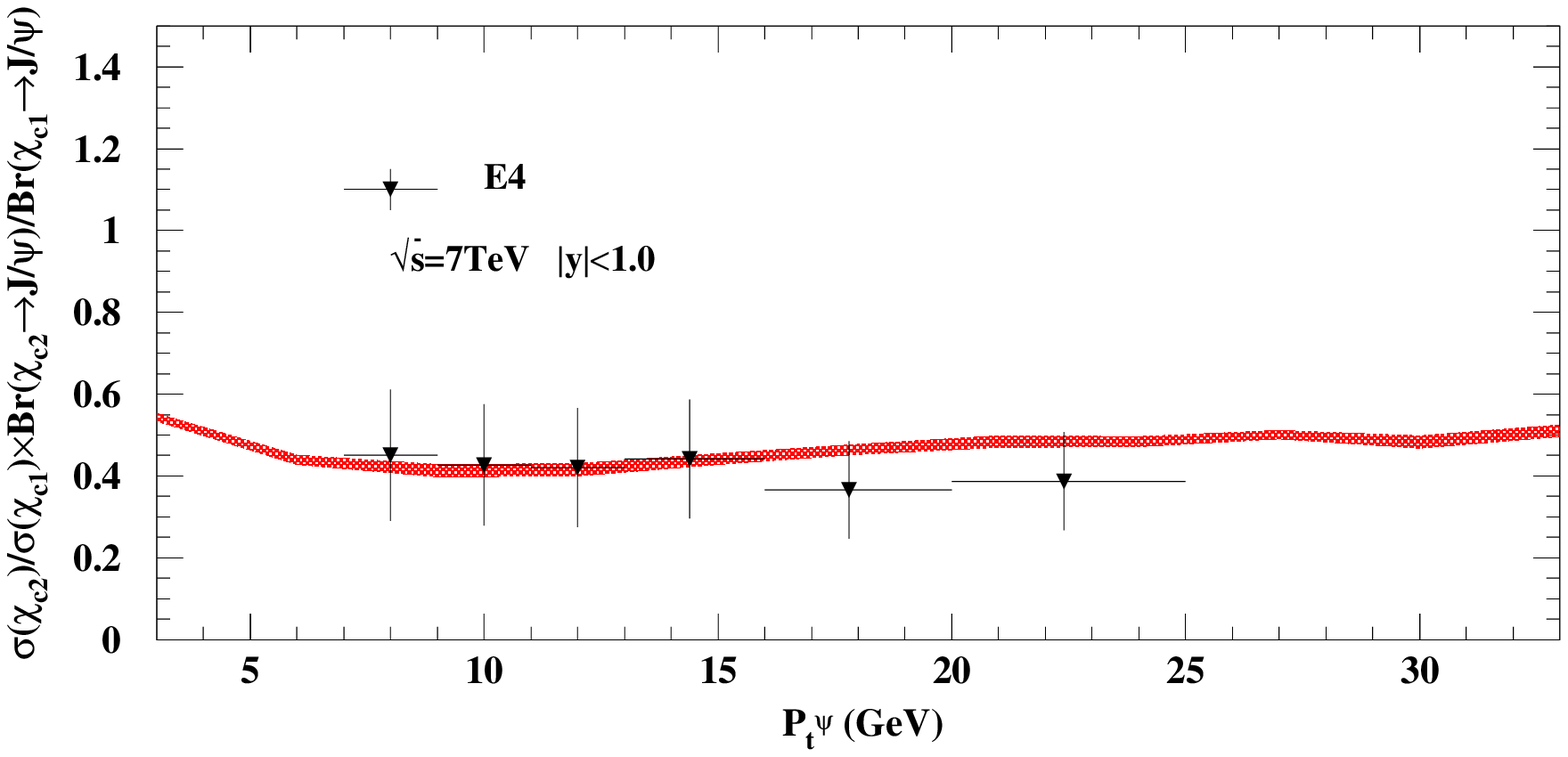}
\caption{\label{fig:nlo} The NLO results for the $\chi_c$ production cross section and the ratio $\sigma(\chi_{c2})/\sigma(\chi_{c1})$
as a function of $p_t$ at the Tevatron and LHC.
The band corresponds to the NLO prediction between the upper and lower bounds
using the eight LDMEs for different renormalization schemes and the values of $\mu_\Lambda$.
The experimental data are taken from Refs.~\cite{Abulencia:2007bra, LHCb:2012af, Aaij:2013dja, LHCb:2012ac, Chatrchyan:2012ub, ATLAS:2014ala}.}
}
\end{figure}

We also present here the LDME obtained with the inclusion of E2 and E3, and see whether the results change much:
\bea
{\cal O}_{\mu_\Lambda}^{NLO}=2.09\pm 0.04,~~~~~~{\cal O}_{m_c}^{NLO}=2.35\pm 0.04, \NO \\
{\cal O}_{m_c/2}^{NLO}=1.77\pm 0.04,~~~~~~{\cal O}_{\Lambda_{QCD}}^{NLO}=0.77\pm 0.04, \label{eqn:LDMEscalecom}
\eea
where the subscript $\mu_\Lambda$ denotes the $\mu_\Lambda$-cutoff renormalization scheme as well as $\mu_\Lambda=m_c$,
and the subscripts $m_c$, $m_c/2$ and $\Lambda_{QCD}$ refer to the $\overline{MS}$ renormalization scheme, with $\mu_\Lambda$ being the corresponding values.
The $\chi^2/d.o.f.$ are 0.51, 0.52, 0.50, and 0.46, respectively.
The difference between the LDMEs fitted by including and excluding E2 and E3 ranges from $4\%$ to $10\%$.
The difference increases as $\mu_\Lambda$ gets smaller,
which is caused by the different behavior of $r$ for the four experimental conditions.
Including E2 and E3 enhances the $\chi^2/d.o.f.$ slightly,
which is to say that theoretical prediction can fit E2 and E3 equally as well as it fits E1 and E4$-$E6.
Figure\ref{fig:nlo} presents the comparison of theoretical predictions for the eight LDMEs to the experimental data.
Actually, all the eight LDMEs result in good agreement with the experiment.
For E1 and E4$-$E6, the bands hold small as $p_t$ varies,
while for E2 and E3, the bands get very large in high $p_t$ regions,
which is to say for E1, E4$-$E6, and the small $p_t$ region in E2 and E3, the $\mu_\Lambda$ dependence can be absorbed into the LDMEs,
while in the large $p_t$ regions in E2 and E3, the problem of $\mu_\Lambda$ dependence becomes severe.

The values of $\langle O^{\chi_{c0}}(^3S_1^{[8]})\rangle$ at LO obtained in this paper, which range from 0.00013GeV$^3$ to 0.00126GeV$^3$,
are smaller than those in Ref.~\cite{Cho:1995ce} ((0.00327$\pm$0.00043)GeV$^3$), Ref.~\cite{Braaten:1999qk} ((0.0019$\pm$0.0002)GeV$^3$),
and Ref.~\cite{Sharma:2012dy} ((0.00187$\pm$0.00025)GeV$^3$),
which employed the data obtained through extrapolation carried out in Ref.~\cite{Abe:1997yz}.
The NLO LDMEs are slightly different from those obtained in Refs.~\cite{Ma:2010vd, Gong:2012ug},
which is due to the different parameter choices between our paper and those cited above.
With the same parameter choice, we can obtain exactly the same LDMEs with Refs.~\cite{Ma:2010vd, Gong:2012ug}.

\section{summary}

In this paper, we calculated $\chi_c$ production cross sections and the ratio $\sigma(\chi_{c2})/\sigma(\chi_{c1})$ at hadron colliders,
and compared the theoretical predictions with the experiment.
We presented a detailed analysis on the CO LDMEs and found that,
at LO, there does not exist any universal value of the CO LDME to explain all the experiments,
while at NLO, the CO LDME obtained from a global fit is able to explain all the experimental data.
At LO, we obtained the value of ${\cal O}$ ranging from 0.13 to 1.26 when fitting individual experiments E1$-$E6.
The upper and lower bounds of ${\cal O}$ result in quite a large band,
which can cover most of the experimental data; however,
the upper bound overestimates the significance of $\chi_c$ feeddown contributions for some of the experimental conditions.
As for the NLO case, we carried out a global fit by using eight schemes.
Each of them agree well with the experimental data.
We also investigated the $\mu_\Lambda$ dependence of the results and found that,
for E1, E4$-$E6, and the small $p_t$ region in E2 and E3, the dependence on $\mu_\Lambda$ can be absorbed into the LDMEs,
while for the large $p_t$ region in E2 and E3, the problem of $\mu_\Lambda$ dependence is relatively severe.
One needs to resum the large log terms rising from large rapidity to achieve better results.
Our work provides a strong support of the NRQCD effective theory.

Acknowledgments.---We thank Jian-Xiong Wang for helpful discussion.
This work is supported by the National Natural Science Foundation of China (Grant No.~11405268).


\end{document}